\documentclass[sn-nature]{sn-jnl}

\usepackage{graphicx}%
\usepackage{multirow}%
\usepackage{amsmath,amssymb,amsfonts}%
\usepackage{amsthm}%
\usepackage{mathrsfs}%
\usepackage[title]{appendix}%
\usepackage{xcolor}%
\usepackage{textcomp}%
\usepackage{manyfoot}%
\usepackage{booktabs}%
\usepackage{mathrsfs}
\usepackage{pdfpages}

\usepackage{caption}
\DeclareCaptionFont{sncap}{\fontsize{8}{9.5}\selectfont}
\DeclareCaptionLabelSeparator{snfigsep}{\hspace{.7em}}
\usepackage{algorithm}%
\usepackage{algorithmicx}%
\usepackage{algpseudocode}%
\usepackage{listings}%
\usepackage{float}

\theoremstyle{thmstyleone}%

\theoremstyle{thmstyletwo}%

\theoremstyle{thmstylethree}%

\setcitestyle{super,open={},close={},comma,sort&compress}

\begin{document}

\title[Article Title]{Non-Hermitian fluctuations enable model-free particle manipulation}

\author*[1]{\fnm{Siarhei} \sur{Zavatski}}\email{siarhei.zavatski@epfl.ch}

\author[2]{\fnm{Tristan} \sur{Nerson}}\email{tristan.nerson@epfl.ch}

\author[2]{\fnm{Romain} \sur{Fleury}}\email{romain.fleury@epfl.ch}

\author*[1]{\fnm{Olivier J.F.} \sur{Martin}}\email{olivier.martin@epfl.ch}

\affil*[1]{\orgdiv{Nanophotonics and Metrology Laboratory}, \orgname{Swiss Federal Institute of Technology Lausanne (EPFL)}, \orgaddress{\street{Station 11}, \city{Lausanne}, \postcode{1015}, \country{Switzerland}}}

\affil[2]{\orgdiv{Laboratory of Wave Engineering}, \orgname{Swiss Federal Institute of Technology Lausanne (EPFL)}, \orgaddress{\street{Station 11}, \city{Lausanne}, \postcode{1015}, \country{Switzerland}}}

\abstract{Contactless manipulation of microscopic matter is central to applications ranging from the isolation of circulating tumor cells in liquid biopsies\cite{nagrath2007isolation, li2015acoustic} to the removal of microplastics from environmental water\cite{abdeljaoued2024efficient,chen2022removal,costa2024echogrid}. Electromagnetic approaches are particularly attractive because fields can be structured within compact microfluidic systems using either light or simple electrode architectures. However,  precise manipulation requires calibrated models of the field distribution and accurate knowledge of the properties of both the object and the surrounding medium, which limits applicability to well-characterized, static systems \cite{riccardi2023electromagnetic}. Here we show that energy dissipation itself provides sufficient information for deterministic particle control. Instead of relying on explicit field calibration, our approach exploits an original relationship between particle position, energy dissipation, and electromagnetic body forces, which can be accessed experimentally through variations of conductance matrices. By extracting force-shaping voltage patterns from these measurements, we demonstrate fully automated closed-loop manipulation of silica microbeads in one and two dimensions, including in the presence of other freely moving particles in a disordered background. These results establish a pathway toward deterministic force control by deliberately measuring and exploiting the non-Hermitian response of the system to engineer electromagnetic momentum transfer. This framework expands micromanipulation into realistic, dynamically evolving environments, where wave-matter interactions cannot be fully pre-characterized or eliminated through design.}

\keywords{electromagnetic force, micromanipulation, dielectrophoresis, model-free, non-Hermitian, loss, reconfigurable, microfluidics, lab-on-a-chip}

\maketitle

Among existing particle micromanipulation methods\cite{cha2022multiphysics, schmidt2025three, cai2023magnetic, katzmeier2023microrobots, ozcelik2018acoustic, yang2021optical, zhang2021plasmonic, qin2026nanoscale, zhan2023versatile}, electromagnetic approaches, including dielectrophoresis (DEP)\cite{pethig2017dielectrophoresis, tian2024chip}, optoelectronic\cite{chiou2005massively, zhang2022optoelectronic}, and optical tweezers\cite{yang2021optical, otte2020optical},  have emerged as particularly versatile tools, capable of exerting controlled forces on particles across a wide range of sizes, compositions, and physical properties. By spatially varying the amplitude, phase, and frequency of applied electromagnetic fields, they can generate advanced force landscapes for trapping\cite{shen2024acousto, sun2019continuous}, sorting\cite{macdonald2003microfluidic, farasat2022signal}, transporting\cite{jia2024precise,yang2023optofluidic,chen2026programmable}, and assembling\cite{he2026flexible,li2023precise} microscale particles.  

Despite these capabilities, all programmable electromagnetic manipulation schemes remain fundamentally constrained by the need for an explicit model of the particle–field interaction to predict the forces arising in a given experimental setting.
Actuation signals that steer particles along desired paths are either determined via real-time numerical simulations\cite{lefevre2022closed,lefevre2024real,lefevre2023automatic,zemanek2018phase,julius2023dynamic,julius2024portable,erben2025optical} or programmed in advance using prior knowledge of the system\cite{liu2022pomdp,huang2021optimization,zheng2025automated,zaman2022controlled}. Consequently, the manipulation procedure becomes inherently system-specific, and even small parameter fluctuations or incomplete knowledge of system properties can result in significant manipulation errors. Although feedback-controlled systems\cite{lefevre2022closed,zemanek2018phase,zemanek2015feedback,liu2022pomdp} can relax the accuracy requirements imposed on spatial field models, they remain strongly dependent on precise knowledge of the dielectric properties of both the particle and the surrounding medium, e.g., the Clausius–Mossotti factor\cite{pethig2026clausius, pethig2019limitations}. Conversely, machine-learning-based force prediction approaches can bypass explicit force models altogether\cite{fang2023data,lenton2020machine,ajala2021deep,su2021machine}, yet they still depend on training data that often fail to generalize across different particle types and environments. As a result, the broader deployment of these methods beyond laboratory settings remains limited, since most real-world samples consist of complex and evolving mixtures of particulates with diverse compositions and morphologies, making comprehensive {\em a priori} characterization impractical.

To overcome these challenges, wavefront-shaping techniques have recently been proposed for deterministic particle control  without requiring any parametric model based on the particle's material properties or geometry\cite{ambichl2017focusing,cao2022shaping,butaite2024photon,horodynski2023tractor,orazbayev2024wave,nerson2025optimal,horodynski2020optimal,liu2018optimal,ren2022rigorous,rakich2009general}. These methods retrieve the scattering matrix from far-field measurements of a linear system, thereby providing complete information about its local structure, physical properties, and even its dynamical evolution\cite{hupfl2023optimal,hupfl2023optimal2}. Despite the promise of wavefront shaping for system-agnostic particle control, experimental progress has only been confined to two isolated demonstrations in acoustics\cite{orazbayev2024wave} and in the microwave electromagnetic regime\cite{horodynski2020optimal}. In acoustics, wavefront shaping has enabled arbitrary translational and rotational positioning of macroscopic particles (ping-pong balls), albeit so far on practically irrelevant timescales of several hours. The microwave experiment has shown that far-field wavefronts can be engineered to produce near-field patterns capable of transferring momentum to a particle, yet without demonstrating actual particle movement. A fundamental obstacle underlies both cases: wavefront shaping frameworks assume a lossless system, ensuring that no wave-carried momentum is dissipated between the far-field ports and the particle. Real-world microfluidic environments, however, strongly violate this assumption, as ionic solutions, lossy disordered media, and substrates all introduce absorption. Overcoming this limitation would extend  electromagnetic force shaping to the broad class of systems that underpin practical micromanipulations, from point-of-care diagnostics to single-cell biology. Unfortunately, no efforts have yet been dedicated to extending wavefront shaping theory for particle manipulation beyond Hermitian systems, for example into the electroquasistatic regime that is routinely applied in microfluidic settings.

Here, we address this question by developing dissipation-guided electromagnetic manipulation (DGEM), a method that extends wavefront shaping principles to quasistatic, dissipative systems. The core component of DGEM is the variational conductance (VC) operator, evaluated at the electrode boundaries, which reveals the spatial distribution of energy dissipation. Crucially, rather than being hindered by the breaking of Hermiticity due to losses, the VC operator explicitly exploits non-Hermitian information: conductance perturbations induced by a particle exactly encode the local DEP forces, enabling the optimal actuation voltage pattern to be identified via solving a simple eigenvalue problem without prior calibration or assumptions about the system’s composition.
To demonstrate DGEM performance, we conduct several experiments using a microfluidic chip comprising an array of 20 individually addressable planar metal electrodes, achieving successful one- and two-dimensional manipulation of 20 $\mu$m silica microbeads -- including in the presence of other freely moving particles -- with particle translation velocities of about 7.5~$\mu$m/s and with fully closed-loop automated control. Remarkably, the theoretical foundation of the proposed method supports its extension to manipulate a wide range of target particles -- including polymer particles, biological cells, extracellular vesicles, microdroplets, and arbitrarily shaped conductive particles -- opening a practical pathway toward automated precision cell handling, droplet microfluidics, targeted drug delivery, and single-cell studies.

\section*{Working principle}\label{sec2}

\begin{figure}[tp!]
\centering
\includegraphics[width=1\textwidth]{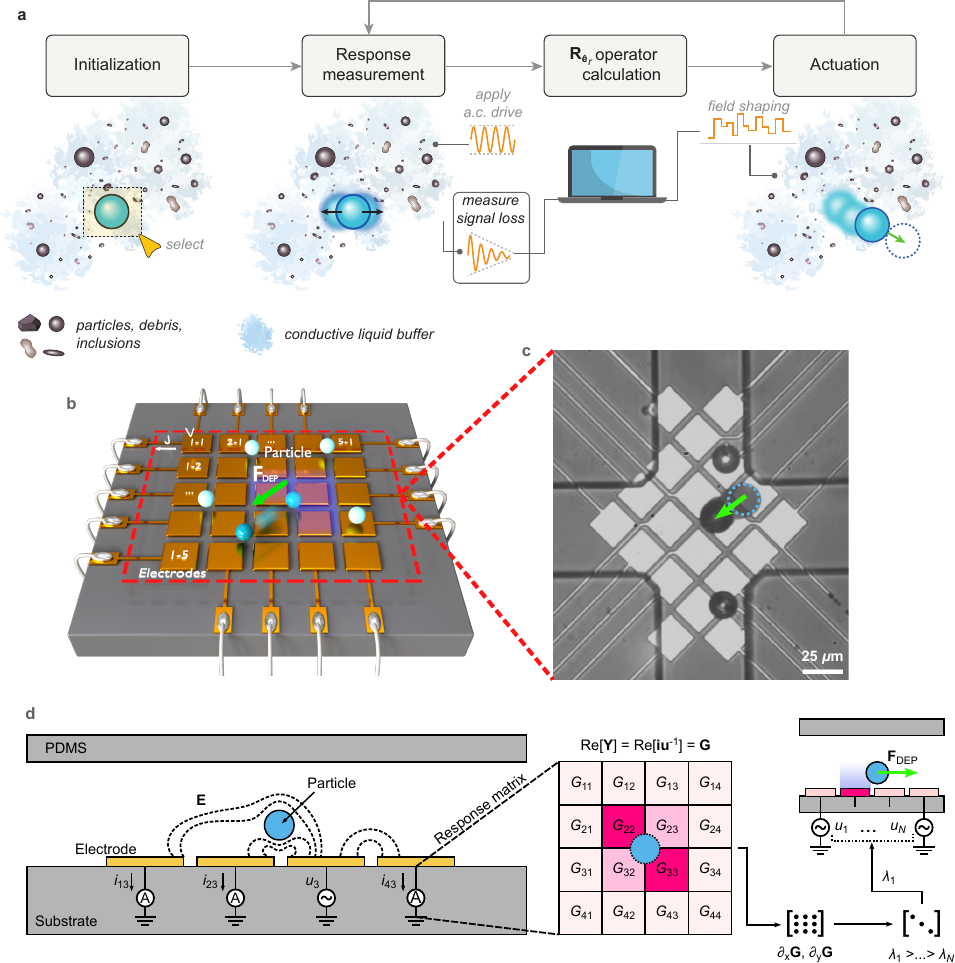}
\caption{\textbf{Principle of dissipation-guided electromagnetic manipulation.} (a) Schematic illustration of the sequence of steps used to perform particle manipulation within a heterogeneous conductive liquid buffer containing multiple constituents. (b) Schematic of the boundary electrodes used to measure the conductance matrix $\mathbf{G}$ and apply voltage patterns that drive particle motion. (c) Optical image of the manipulation experiment demonstrating DGEM-driven particle translation. (d) Cross-sectional illustration of interaction between the particle and the electric field generated  by the microelectrodes. The presence of the particle perturbs the dissipated power, which is detected through changes in the conductance matrix $\mathbf{G}$, as illustrated schematically on the right. By measuring $\mathbf{G}$ at different particle positions, its spatial derivative is computed and used to construct the VC operator. Diagonalization of this operator yields eigenvectors $\mathbf{u}$ and eigenvalues $\mathbf{\lambda}$. Selecting the eigenvalue with the largest magnitude ensures that the corresponding input signal pattern, proportional to $\mathbf{u}$, maximizes the DEP force acting on the particle.}
\label{f1}
\end{figure}

Figure~\ref{f1} summarizes the working principle of the DGEM method and its application to quasistatic electric fields in dissipative, inhomogeneous systems. The detailed control logic is outlined in Extended Data Fig.~\ref{ED1}, and the experimental setup and fabrication procedures are described in the Methods section. A typical manipulation sequence proceeds in four steps: initialization, response measurement, operator calculation, and actuation (Figure~\ref{f1}a). First, a conductive liquid buffer containing 20~$\mu$m-diameter silica microbeads is introduced into the microfluidic chamber and positioned between or above the metallic electrode array (Figure~\ref{f1}b-c). The target particle is then selected and displaced in any direction using a random actuation signal, bringing the system to a new equilibrium state. The electrical response of this state -- specifically, the conductance matrix $\mathbf{G}$, which encodes the power dissipation -- is subsequently measured. Repeating these measurements at multiple particle positions and differentiating the resulting matrices to obtain their spatial derivatives (see Methods and Extended Data Fig.~\ref{ED7} for the detailed matrix acquisition process), we are able to link the particle-induced reshaping of the electric field and current distribution to electromagnetic forces.

The theoretical foundation of this procedure is a port-based force identity that relates small changes in the measured electrical response to mechanical work. In the Supplementary Information, we introduce this technique by first deriving the time-averaged electromagnetic force density in inhomogeneous, lossy, isotropic media from the complex Lorentz force law, building on microscopic formulations of electromagnetic force densities\cite{griffiths2015s,anghinoni2023microscopic, kaiser2014completing,nieto2022complex}. For a rigid object immersed in an incompressible fluid, the force density can be decomposed into a non-gradient component that drives rigid-body motion, and a pure-gradient component that is absorbed into the hydrodynamic pressure and produces no net force or torque. We further show that the non-gradient force can be equivalently obtained from the Minkowski stress tensor evaluated on a closed surface in the homogeneous exterior medium, including in lossy media (Supplementary Note 1). Crucially, this force can be accessed from the admittance matrix measured at the electrodes. For a voltage vector $\mathbf u$, a small rigid displacement $\delta \mathbf r$ of the target particle induces a change in the admittance matrix $\mathbf Y=\mathbf G+j\mathbf B$. If the system could be considered as lossless, we would be able to estimate the mechanical work from the susceptance variation
\begin{equation}
\mathbf F\cdot \delta \mathbf r = -\frac{1}{4\omega}\mathbf u^\dagger \delta \mathbf B\,\mathbf u.
\end{equation}

This lossless scenario represents the counterpart of identities based on Hermitian response operators \cite{nerson2025optimal} (Supplementary Note 2). However, this strategy is practically irrelevant in the non-Hermitian cases considered here, due to the non-negligible conductivity in the medium and particles. Instead, DGEM relies on the fact that, surprisingly, a similar estimation principle can be devised in lossy conductive systems, but this time based on variations of the conductance matrix $\mathbf G$, which encodes non-Hermitian fluctuations. Let $\varepsilon(\mathbf r)$ and $\sigma(\mathbf r)$ denote the permittivity and Ohmic conductivity distributions of the physical system. When $\omega\varepsilon(\mathbf r)\ll\sigma(\mathbf r)$ in each material domain, the electric potential $\phi(\mathbf r)$ is governed by a real conduction problem $\nabla\cdot[\sigma(\mathbf r)\nabla \phi(\mathbf r)]=0$. Remarkably, this non-Hermitian problem can be exactly mapped onto an auxiliary lossless dielectric problem with $\sigma^\sharp(\mathbf r)=0$ and $\varepsilon^\sharp(\mathbf r)=\sigma(\mathbf r)/\omega$, whose susceptance matrix matches the limiting conductance matrix of the physical system. Comparing the physical and auxiliary stresses in the exterior medium gives the low-frequency conductance-force relation
\begin{equation}\label{equivalence}
\mathbf F\cdot \delta \mathbf r = -\frac{\varepsilon_m}{4\sigma_m}\mathbf u^\dagger \delta \mathbf G\,\mathbf u ,
\end{equation}
where $\varepsilon_m$ and $\sigma_m$ are the permittivity and conductivity of the liquid buffer in contact with the electrodes (Supplementary Note 3). While numerical simulations show that $\mathbf{B}$-matrices can sometimes yield eigenstates producing force directions correlated with the target direction, this agreement is imperfect and significantly less robust than for $\mathbf{G}$. In the low-frequency, conduction-dominated regime considered here, $\mathbf{G}$ provides an exact force retrieval, both in magnitude and direction, making conductance the natural observable for DGEM (Supplementary Note 4). Experimentally, conductance also yields a higher signal-to-noise ratio (SNR) than susceptance measurements, consistent with the Ohmic response of the ionic aqueous buffer.

Equation \eqref{equivalence} establishes the equivalence principle underlying DGEM: the mechanical information contained in the reactive response of an auxiliary lossless dielectric system is mapped, in a conductive low-frequency system, onto the dissipative response measured via conductance. DGEM therefore only requires measuring how the conductance matrix changes when the particle is displaced. For a prescribed direction $\hat{\mathbf e}_r=\delta\mathbf r/|\delta\mathbf r|$, we define the VC operator
\begin{equation}\label{eq:R_er}
\mathbf R_{\hat{\mathbf e}_r} = -\frac{\varepsilon_m}{4\sigma_m}\hat{\mathbf e}_r\cdot\nabla_{\delta\mathbf r}\mathbf G ,
\end{equation}
so that
\begin{equation}
\mathbf F\cdot \hat{\mathbf e}_r = \mathbf u^\dagger \mathbf R_{\hat{\mathbf e}_r}\mathbf u .
\end{equation}
Diagonalizing $\mathbf R_{\hat{\mathbf e}_r}$ directly gives the electrode voltage patterns that extremize the force along $\hat{\mathbf e}_r$: the eigenvalue fixes the force magnitude and sign, while the corresponding eigenvector gives the actuation pattern. Iterating this cycle of measurement, operator construction, and actuation as the particle moves closes the loop, enabling fully autonomous manipulation.

\section*{Linear translation in one dimension}\label{subsec2.2}

\begin{figure}[tp!]
\centering
\includegraphics[width=1\textwidth]{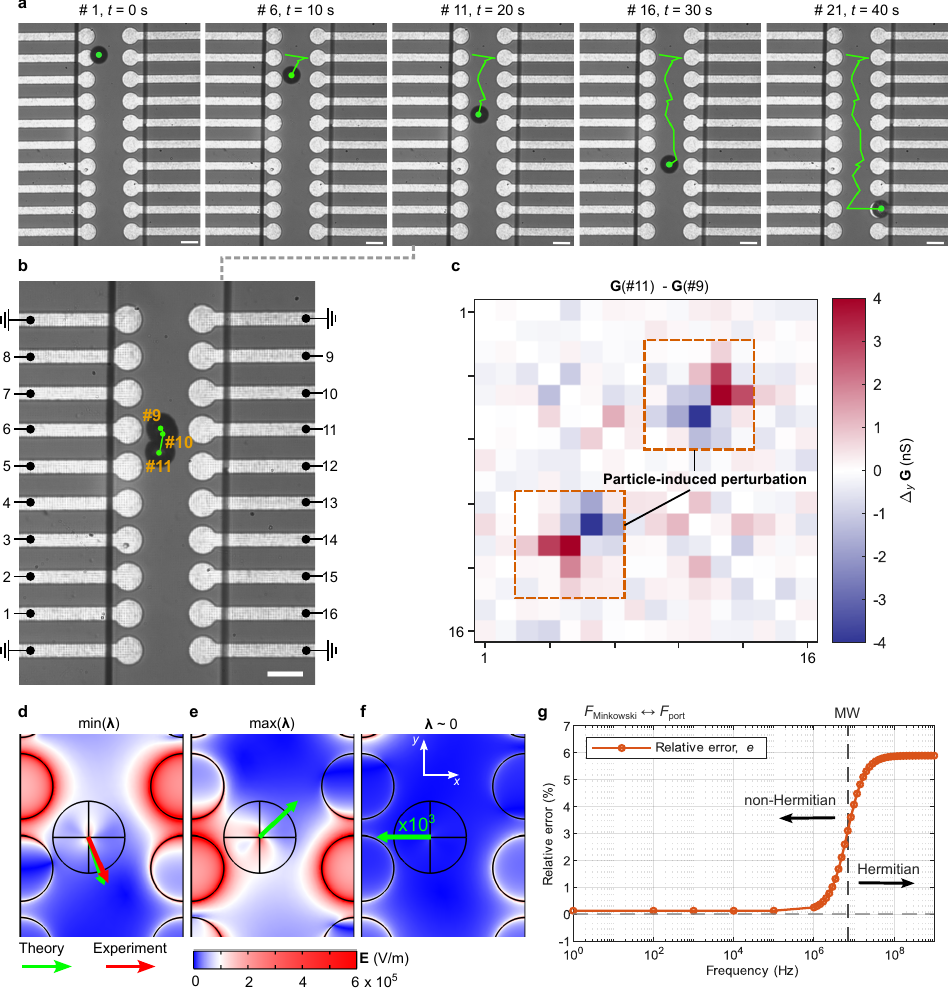}
\caption{\textbf{Closed-loop particle manipulation from conductance variations}. (a) Sequential microscopy images showing one-dimensional motion of a 20 $\mu$m silica particle. Images were acquired at different iterations, and the particle trajectory is indicated by the green line. (b) Overlay of 3 microscopy images acquired at iterations \#~9, \#~10, and \#~11. Numbers indicate the electrode addressing sequence used for admittance matrix construction. (c) Differential conductance matrix computed via finite differences using measurements from iterations \#~9 and \#~11, shown in (b). (d-f) Simulated electric field distributions near the particle when electrodes are driven with eigenvectors corresponding to the (d) minimum, (e) maximum, and (f) near-zero eigenvalue $\lambda$. The green arrow indicates the force direction calculated from the Minkowski stress tensor, while the red arrow denotes the experimentally observed particle displacement after iteration \#~11. (g) Theoretical frequency dependence of the relative error, $e = 100 \% \cdot |(F_{\text{Minkowski}} - F_{\text{port}})/F_{\text{Minkowski}}|$, between the $y$-component of the force computed using the Minkowski stress tensor, $F_{\text{Minkowski}}$, and that predicted by the VC operator, $F_{\text{port}}$, below the Maxwell-Wagner (MW) relaxation frequency. Simulations use experimentally measured conductivity for the surrounding medium, while the effective particle conductivity is estimated using reference data\cite{weng2016size,leroy2013influence}. All scale bars are 25~$\mu$m.}
\label{f2}
\end{figure}

We first demonstrate DGEM performance by manipulating a single particle in one dimension. Sequential microscopy images in Figure~\ref{f2}a and Supplementary Video~1 show the movement of a 20 $\mu$m-diameter silica microbead from the upper to the lower boundary of the experimental system over 20 iterations, covering approximately 300 $\mu$m in about 40~s. 
This translation speed is governed by the conductance matrix update rate per iteration, which is approximately 2~s in this case. The example of raw electrical signals acquired from all 20 electrodes during conductance matrix measurements is shown in Extended Data Fig.~\ref{ED2}. Under the present experimental conditions, readout of a single conductance element requires approximately 6~ms. The particle-induced conductance perturbation typically ranges from 2 to 4~nS, with absolute levels of 100 to 500~nS and noise levels of 0.1 to 0.5~nS (standard deviation). The achieved manipulation speed is comparable to model-dependent electromagnetic approaches such as DEP\cite{lefevre2023automatic,zaman2022controlled}, as well as optical and optoelectronic tweezers\cite{melzer2018fundamental,zhang2019optoelectronic}, and exceeds by orders of magnitude the relative speed of the only model-free wavefront shaping experiment demonstrated to date, performed in the acoustic domain \cite{orazbayev2024wave}.

Figure~\ref{f2}b overlays three optical images captured at consecutive iterations during particle manipulation and illustrates the electrode addressing sequence used for conductance matrix measurements. Of the twenty available electrodes, sixteen are used for matrix construction and subsequent actuation, while the remaining four boundary electrodes (two at each end) are grounded. We construct a 16$\times$16 conductance matrix from pairwise electrode measurements, with diagonal elements set to zero (Figure~\ref{f2}c; additional examples of the experimental conductance matrices are shown in Extended Data Fig.~\ref{ED3}). Figures~\ref{f2}d–f illustrate how DGEM shapes the electric field in the experimental configuration of Figure~\ref{f2}b when the electrodes are driven with eigenvectors corresponding to the minimum, near-zero, or maximum eigenvalue $\lambda$ of the VC operator. For the minimum $\lambda$ (Figure~\ref{f2}d), the electric field localizes above the particle, producing a DEP force of -164~pN, consistent with the downward translation observed experimentally (compare the red and green arrows in Figure~\ref{f2}d). For the maximum $\lambda$ (Figure~\ref{f2}e), the field localizes below the particle, reversing the force direction and changing its magnitude to 257~pN. Additional simulation results in Extended Data Fig.~\ref{ED4} confirm strong agreement between the sign of the eigenvalue and the direction of the corresponding DEP force. In the special case of a near-zero eigenvalue (Figure~\ref{f2}f), the simulations indicate that the electric field effectively bypasses the particle and the DEP force is reduced by approximately three orders of magnitude. 

Consistent with this interpretation, Figure~\ref{f2}g confirms numerically that, below the Maxwell--Wagner relaxation frequency\cite{pethig2017book}, the force predicted from the VC operator quantitatively agrees with the force obtained from the Minkowski stress tensor using the particle and medium parameter estimates. This confirms that DGEM provides the full set of manipulation modes required for autonomous, arbitrary particle control, including optimal translation in any chosen direction and stable particle trapping. 

\section*{Complex manipulations in microfluidics}\label{subsec2.4}

One geometric parameter that directly influences the SNR, and therefore the manipulation accuracy in our one-dimensional configuration (Figure~\ref{f2}a), is the gap between adjacent electrodes. Numerical simulations in Extended Data Fig.~\ref{Avsg} show that the conductance magnitude of the most sensitive matrix element, measured in the presence of a 20~$\mu$m particle, increases from 0.6 to 27~nS when the electrode gap is decreased from 90 to 30~$\mu$m. Smaller gaps therefore improve electrical sensitivity. This highlights an important design advantage of DGEM: because the method relies on measured conductance variations rather than on a specific electrode geometry, it can be adapted to different layouts depending on the desired balance between sensitivity, channel dimensions, and manipulation range. We next exploit this flexibility in more complex microfluidic configurations, using T-junction electrode geometries for particle routing and, for two-dimensional steering, an electrode array positioned underneath the particle to preserve high electrical sensitivity over a larger manipulation area.

\begin{figure}[tp!]
\centering
\includegraphics[width=1\textwidth]{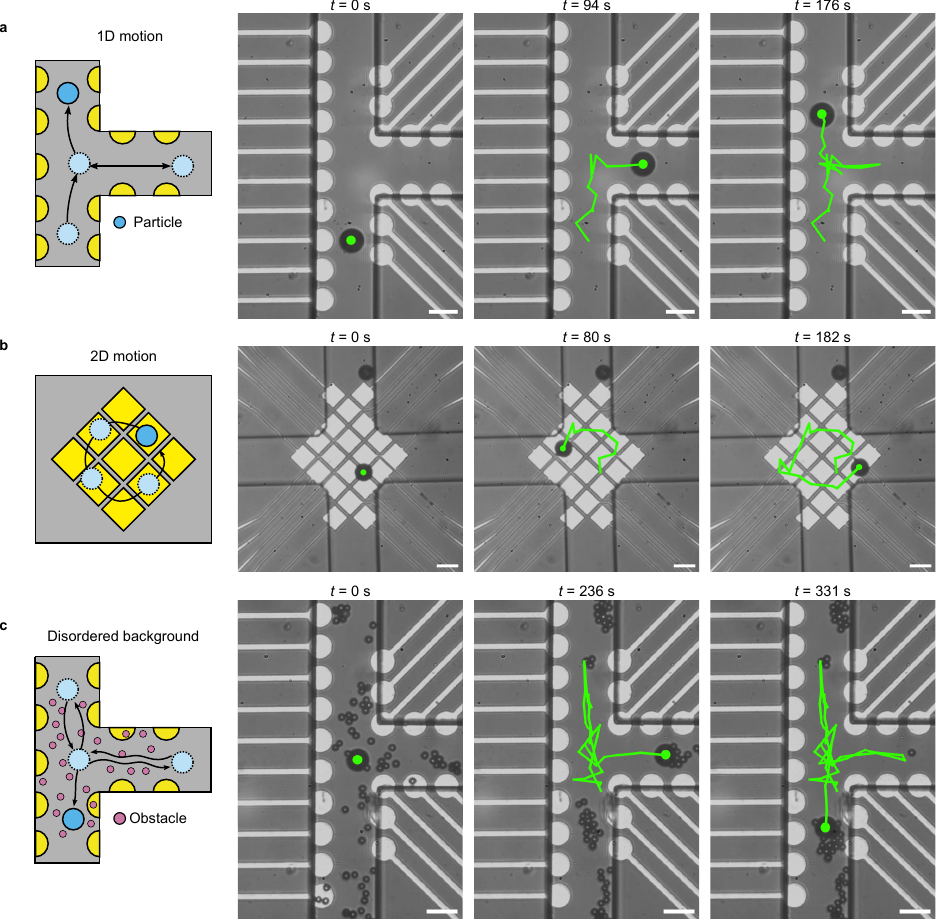}
\caption{\textbf{System-agnostic particle manipulation}. Schematic illustrations and sequential microscopy images demonstrating (a) one-dimensional motion of a silica particle along the $x$- and $y$-directions, (b) two-dimensional particle manipulation, and (c) one-dimensional transport of a 20~$\mu$m silica particle through a disordered background of 5~$\mu$m silica particles. All scale bars are 25~$\mu$m.}
\label{f4}
\end{figure}

The potential of DGEM as a new manipulation paradigm for real-world applications is demonstrated in Figure~\ref{f4}. Using aqueous solutions of 20 $\mu$m diameter silica microbeads, we achieve (i) one-dimensional single-particle motion along both $x$- and $y$-directions (Figure~\ref{f4}a and Supplementary Video~2); (ii) two-dimensional single-particle motion (Figure~\ref{f4}b and Supplementary Video~3); and (iii) one-dimensional single-particle motion in the presence of a dense population of 5~$\mu$m silica particles (Figure~\ref{f4}c and Supplementary Video~4). In all cases, the target particle faithfully follows the DGEM-prescribed translation commands. For example, particles can be steered between microfluidic channel branches, enabling their isolation for downstream processing outside the main collection, as shown in Figures~\ref{f4}a and~\ref{f4}b. Moreover, Figure~\ref{f4}c demonstrates that the selected particle can follow a prescribed trajectory while traversing a crowded environment of uncontrolled particles. In this scenario, the additional particles are free to move during electrode actuation and are not explicitly tracked by the algorithm. Interestingly, due to dipole-dipole interactions\cite{hu1994observation,katzmeier2022emergence,boymelgreen2018active}, the smaller particles tend to form chains and can attach to the target particle under the a.c.\ electric field (see Supplementary Video 4). Once 5~$\mu$m beads adhere to the target particle, its effective shape changes, and it can no longer be treated as an isolated sphere. The position and number of attached particles also evolve during manipulation, leading to time-dependent variations in the target particle geometry. Remarkably, despite these changes, the selected particle continues to follow the prescribed trajectory, in sharp contrast to existing manipulation algorithms that require precise knowledge of the position and shape of all involved particles to infer appropriate actuation signals \cite{liu2022pomdp,zheng2025automated}. This robustness to spatial and temporal drift in system parameters could be particularly important for studying cell interactions under physiological conditions\cite{armingol2024diversification,shen2024acousto,dura2016longitudinal}, cell patterning\cite{ma2020acoustic}, cell surgery\cite{zhang2022microfluidic,shakoor2022advanced}, engineering of micro-swimmers and the exploration of their behavior in complex fluids\cite{bechinger2016active,palagi2016structured}, as well as the fabrication of reconfigurable, smart materials \cite{trivedi2022self,al2020magnetic,yigit2019programmable,han2023electric,barros2025structure}, among other applications.

\section*{Discussion and outlook}\label{sec3}

The results presented here constitute, to our knowledge, the first demonstration of model-free particle micromanipulation in dissipative media using the extended principles of electromagnetic wavefront shaping. Although this framework has so far been established for Hermitian systems, our theoretical and experimental findings indicate that dissipation can be beneficial and, in certain parameter regimes, is in fact essential for effective particle manipulation. These losses, quantified through particle-induced conductance perturbations, directly encode the complex quasistatic electric field patterns that generate DEP forces, enabling guided transport of arbitrarily selected microparticles through unknown environments.

DGEM also differs from conventional electromagnetic manipulation strategies, which rely either on physical models of the system to simulate forces and generate actuation signals, or on heuristic selection of actuation patterns. Instead, it exploits system-specific information encoded in electrical signals that are continuously updated at electrode boundaries, together with particle-position guidestars. As a result, it can be seamlessly integrated into arbitrary microfluidic systems, enabling fully automated operation in complex, dynamically evolving environments without prior calibration or system-specific tuning. 

While our demonstration uses silica microbeads as test objects, the approach is expected to be applicable to a wide range of particles, including polymer or conductive particles of arbitrary shape, biological cells, extracellular vesicles, microdroplets, and their mixtures. It can also accommodate arbitrary numbers of electrodes, geometries, and layouts, with these design choices determining manipulation speed and transport distance. This versatility enables broad applications, including precision cell handling, droplet microfluidics, targeted drug delivery, cell signaling studies, and long-term single-cell investigations, among others. In principle, the method may even be extended to the nanoscale, provided that the induced signal variations exceed the noise floor and that the resulting forces are sufficient for deterministic nanoparticle transport. Finally, the demonstrated link between measured electrical signals and local electromagnetic forces provides a promising avenue for extending the approach toward label-free characterization of particle electromechanical and electrophysiological properties.


\section*{Methods}\label{sec4}
\subsection*{Materials}\label{sec4.1}

SiO$_2$ microbeads (powder) with 5 and 20~$\mu$m mean diameters were purchased from Thermo Scientific. Borate-buffered saline (0.5~M in H$_2$O) was purchased from abcr GmbH. Sodium hydroxide (NaOH, 99~\%) and polyethylene glycol sorbitan monolaurate (TWEEN\textsuperscript{\textregistered} 20, $\geq$40~\% lauric acid) were purchased from Sigma-Aldrich. All chemicals were used as purchased without further purification. 

\subsection*{Fabrication of planar metal microelectrodes adjacent to the microfluidic layer}\label{sec4.2}

A 100 mm oxidized silicon wafer ($\langle100\rangle$, p--type, 1 $\Omega \cdot$ cm, wet oxide of 1~$\mu$m) was first cleaned in a mixture of hot concentrated sulfuric acid and hydrogen peroxide (1:3 piranha solution at 100~$^\circ$C, UFT piranha wet bench) for 10 min followed by treatment in high-frequency oxygen plasma (Tepla GiGabatch) at 1000~W for 3 min with a 400~sccm O$_2$ flow. Next,  electron-beam evaporation was used to deposit 5~nm Ti and 100~nm Au electrode materials. A positive photoresist AZ ECI 3007 was further spin-coated at 6000~rpm for 40 s to produce a 0.6~$\mu$m thick layer. The layer was further baked at 100~$^\circ$C for 90~s. Photoresist was exposed using \textit{i}-line photoresist laser writer (405~nm, Heidelberg Instruments MLA150) with the dose of 115~mJ/cm$^2$. The resist required a post-exposure bake, which was performed in a Süss MicroTec ACS200 at 110~$^\circ$C for 60~s. The photoresist development was completed by dipping the wafer in an AZ 726 MIF solvent for 30~s, followed by thermal reflow at 145~$^\circ$C for 180~s. The patterned substrate was subjected to ion-beam etching (argon source, Veeco Nexus IBE350) at a tilt angle of 10$^\circ$ for 144~s. Immediately after etching, the substrate was cleaned in O$_2$ plasma (Tepla GiGabatch) at 500~W for 120~s to remove a thin layer of photoresist contaminated by argon atoms and redeposited metal. Finally, the photoresist was removed in SVC-14 solution at 70~$^\circ$C, and the substrate with metal electrodes was cleaned once again in O$_2$ plasma (Tepla GiGabatch) at 200~W for 1~min to ensure complete removal of organic debris.

\subsection*{Fabrication of SU-8 mold}\label{sec4.3}

A 100 mm silicon wafer ($\langle100\rangle$, p--type, 1 $\Omega \cdot$ cm) was first cleaned in a mixture of hot concentrated sulfuric acid and hydrogen peroxide (1:3 piranha solution at 100~$^\circ$C, UFT piranha wet bench) for 10 min followed by treatment in high-frequency oxygen plasma (Tepla GiGabatch) at 1000~W for 3 min with a 400~sccm O$_2$ flow. Next, a structural photoresist, SU-8 (Kayaku 3050), was spin-coated (Sawatec LSM-250) to a thickness of 40~$\mu$m. The layer was further baked at 95~$^\circ$C for 900~s (Sawatec HP-200). SU-8 exposure was performed using \textit{i}-line photoresist laser writer (375~nm, Heidelberg Instruments MLA150) with the dose of 2500~mJ/cm$^2$. Immediately after this, SU-8 was subjected to a post-exposure bake at 65~$^\circ$C for 60~s followed by an additional bake at 95~$^\circ$C for 180~s (Sawatec HP-200). SU-8 was developed in propylene glycol monomethyl ether acetate (PGMEA, $\geq$99.5~\%) for 180~s, followed by an isopropyl alcohol (IPA, $\geq$99.5~\%) rinse for 60~s and drying under nitrogen flow. Finally, the SU-8 pattern was hard baked at 135~$^\circ$C for 2~h (Sawatec HP-200).

\subsection*{Fabrication of the microfluidic layer}\label{sec4.4}

The fabricated SU-8 mold was used to pattern a polydimethylsiloxane (PDMS, Sylgard\textsuperscript{\textregistered} 184) layer. The fabrication process of the PDMS microfluidic layer started with surface conditioning of the SU-8 mold for 20~min in a vacuum chamber containing 1~mL of chlorotrimethylsilane (TMCS, $\geq$ 99~\%) dispensed in a small vial. Next, PDMS was prepared by thorough mixing of 33~g of fresh Sylgard\textsuperscript{\textregistered} 184 with 3~g of Sylgard\textsuperscript{\textregistered} cross-linker. The prepared mixture was degassed in a vacuum desiccator to remove bubbles. After the bubbles were removed, the mixture was poured onto the SU-8 mold to form a layer approximately 1~cm thick. The viscous mixture above the mold was further cured in a thermostat at 80~$^\circ$C for about 24~h, followed by demolding and punching of solid PDMS to finalize the microfluidic layer. To bond the free-standing PDMS layer and the substrate with electrode arrays, the later was initially activated by O$_2$ plasma at 29~W for 45~s (Plasma HARRICK). Immediately after activation, the microfluidic layer was aligned with the substrate under  microscope and brought into contact. Bonding was completed by thermal annealing of the structure at 80~$^\circ$C overnight.

\subsection*{Experimental setup}\label{sec4.5}

Extended Data Fig.~\ref{ED5} provides a schematic overview of the DGEM experimental setup. Both actuation and conductance measurements were performed using a lock-in amplifier (HF2LI, Zurich Instruments, sampling at 28.78~kSa/s rate, 85~$\mu$s low-pass filter bandwidth) and a transimpedance amplifier (TIA, HF2TA, Zurich Instruments, 100~k$\Omega$ gain) coupled to a programmable electrode-addressing stage implemented on a custom-designed printed circuit board (PCB). In actuation mode, each electrode was assigned a signal amplitude proportional to the corresponding component of the estimated VC operator eigenvector. To apply these amplitudes to 20 individual electrodes, the base a.c.\ voltage (5~V peak-to-peak at 100~kHz) generated by the lock-in amplifier was divided using digital potentiometers (MCP41U83, Microchip Technology). Extended Data Fig.~\ref{ED6} presents the functional block diagram of the experimental setup, including the PCB controlled by a microprocessor (Arduino Mega 2560) interfaced with MATLAB R2025a. Actuation and measurement modes were configured  using four cross-point switches (ADG2128, Analog Devices), enabling signal routing between the excitation source and the measurement circuit.

Before each manipulation experiment, the microfluidic channels were flushed with 0.5~M NaOH for 5~min, followed by a 10~mM borate buffer containing 0.05~\% Tween 20 for an additional 5~min.  The particles were then dispersed in the same buffer at a concentration of 7.5~mg/mL and introduced into the channels using a syringe. After positioning them above the electrodes, chamber inlets were sealed with scotch tape to prevent liquid evaporation and fluid flow.

Particle motion was monitored in real time using optical microscopy combined with digital video tracking. Images of the manipulation region were continuously acquired, and the particle centroid was identified in each frame using a Python-based tracking routine with manual initialization followed by automatic frame-to-frame updating. The measured particle coordinates were time-stamped and transmitted to the MATLAB control software, which used them to associate each admittance measurement with the corresponding particle position.

\subsection*{Acquisition of conductance matrices}\label{sec4.6}

Extended Data Fig.~\ref{ED7} illustrates the procedure used to acquire the conductance matrix. At each measurement step, the conductance matrix $\mathbf{G}$ was reconstructed from the demodulated terminal responses of the electrode array. To measure an off-diagonal matrix element (e.g., $G_{21}$ in Extended Data Fig.~\ref{ED7}), one electrode in the array (electrode~1) was driven by the sensing signal (0.35~V peak-to-peak at 100~kHz), while all remaining electrodes (electrodes~3 and~4) except one (electrode~2) were grounded. The electrode that was not grounded (electrode~2) was connected to the sensing circuit, enabling measurement of the conductance between it and the driven electrode. To measure a diagonal matrix element (e.g., $G_{11}$), one electrode in the array (electrode~1) was driven by the sensing signal, while all remaining electrodes were connected to the transimpedance amplifier (TIA). This procedure was repeated sequentially for all electrodes until the entire conductance matrix was populated. In most experiments, however, the diagonal elements were omitted because their noise floor frequently exceeded the particle-induced signal of interest. This elevated noise likely originates from the acquisition scheme itself, as the diagonal measurements combine currents from all off-diagonal electrodes and therefore accumulate their associated noise contributions.

\subsection*{Signal postprocessing and operator construction}\label{sec4.7}

Positional gradients of $\mathbf{G}$, operator eigenanalysis, and the construction of optimal actuation signal patterns were performed following a procedure similar to that used in previously reported acoustic experiments\cite{orazbayev2024wave}. Briefly, spatial derivatives were estimated using the finite-difference method, while the actuation signal pattern -- defined by the operator eigenvector corresponding to either the maximum or minimum eigenvalue -- was selected according to the desired direction of particle displacement. In contrast to the acoustic experiments, all electrical drive signals were applied in phase and scaled within the 0--5~V peak-to-peak range. These signals were simultaneously applied to the individual electrodes to generate a non-uniform electric field, thereby producing the DEP force responsible for particle motion.

\subsection*{Simulation of the electric field distribution and DEP force}\label{sec4.8}

Two-dimensional quasistatic electric field simulations were performed using the AC/DC module of COMSOL Multiphysics~6.4. The spatial distribution of the electric field strength was obtained by solving the following equation, $-\nabla\cdot\left[(\sigma+j\omega\varepsilon)\nabla\phi\right]=0$, for a system corresponding to the experimental platform shown in Figure~\ref{f2}, where $\phi$ denotes the electric potential at the frequency $\omega$ and the electric field is given by $\mathbf{E} = -\nabla \phi$. To reproduce the experimental conditions, the electrodes were driven with potentials $\phi$ corresponding to the components of the selected eigenvector. 

All simulations were carried out in a water background with relative permittivity $\varepsilon_m = 78$ and electrical conductivity $\sigma_m = 0.03$~S/m, matching the experimental conditions. The electrodes were modeled as gold with $\sigma_\text{gold} = 45.6 \cdot 10^6$~S/m and $\varepsilon_\text{gold} = 6.9$. A 20 $\mu$m--diameter silica particle was modeled using $\sigma_p = 10^{-4}$~S/m and $\varepsilon_p$ = 3.7. The time-averaged DEP force acting on the particle was calculated via the Minkowski stress tensor approach implemented in COMSOL. All simulations were performed in the frequency domain at 100~kHz, with a minimum mesh size of 1~nm used for spatial discretization.

\section*{Data availability}

The data that support the findings of this study are available from the corresponding authors upon reasonable request.

\section*{Acknowledgements}

The authors acknowledge Georg Fantner for providing the transimpedance amplifier used in the experiments.
T.N.\ and R.F.\ acknowledge the financial support of the Swiss National Science Foundation (SNSF) under the grant No.\ 10001567. S.Z.\ acknowledges the financial support of the Swiss National Science Foundation (SNSF) under the grant No.\ 228838.

\section*{Author information}
\subsection*{Authors and Affiliations}
\textbf{Nanophotonics and Metrology Laboratory, Swiss Federal Institute of Technology Lausanne (EPFL), Lausanne, Switzerland}\hfill \break
Siarhei Zavatski \& Olivier J.F. Martin\hfill \break
\textbf{Laboratory of Wave Engineering, Swiss Federal Institute of Technology Lausanne (EPFL), Lausanne, Switzerland}\hfill \break
Tristan Nerson \& Romain Fleury

\subsection*{Contributions}

S.Z.\ and O.M.\ conceived the idea. T.N.\ and R.F.\ developed the theory. S.Z.\ developed the experimental setup and method, performed fabrication and experiments, conducted numerical simulations, and captured datasets. S.Z.\ and T.N.\ created the figures and wrote the original draft. O.M.\ and R.F.\ supervised all aspects of the project and critically reviewed and edited the manuscript. 

\subsection*{Corresponding author}

Correspondence and requests for materials should be addressed to S.Z. or O.M.

\section*{Ethics declarations}
\subsection*{Competing interests} 

S.Z. and O.M. are inventors on European patent application no. EP26178241.1 submitted by the Swiss Federal Institute of Technology Lausanne (EPFL). The other authors declare no competing interests.

\section*{Supplementary Information}

Supplementary Information is available for this paper.

\newpage


\section*{Extended Data}\label{secA1}

\renewcommand{\figurename}{Extended Data Fig.}
\setcounter{figure}{0} 

\begin{figure}[bp!]
\centering
\includegraphics[width=0.6\textwidth]{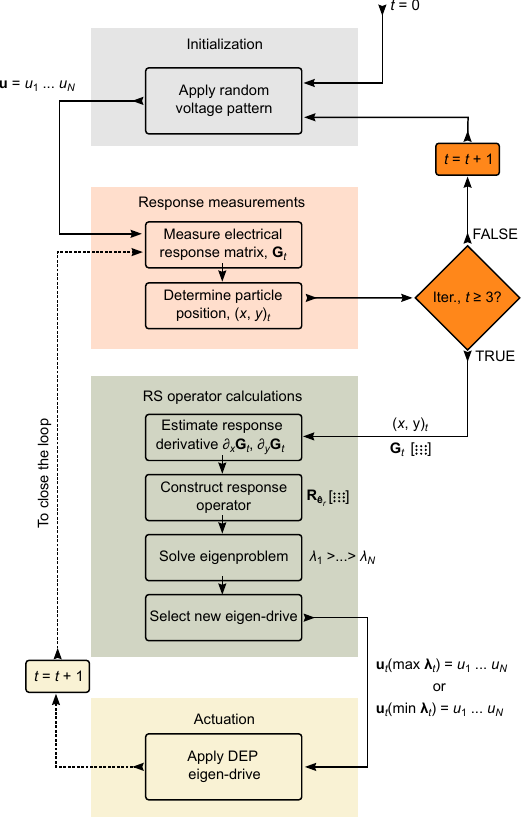}
\caption{\textbf{Closed-loop DGEM control algorithm}. The procedure begins with an initialization stage, during which the particle is displaced using three random electrical signal patterns, $\mathbf{u}$, applied to the electrode array. This stage is performed at the start of the control sequence to acquire the data needed for estimating the spatial gradients of $\mathbf{G}$. Any combination of signal amplitude, frequency, and phase that enables sufficient and safe particle displacement may be used during initialization. After each displacement, the conductance matrix of the system is measured, and the updated  particle coordinates, $x$ and $y$, are recorded. The initialization signal patterns are selected manually by the operator through the control software. Once sufficient data have been collected, DGEM constructs the VC operator, $\mathbf R_{\hat{\mathbf e}_r}$, using Eq.~\ref{eq:R_er}. The operator is then diagonalized to obtain eigenvalues, $\mathbf{\lambda}$, and the corresponding eigenvectors, $\mathbf{u}$. Depending on the desired direction of the next displacement, the eigenvector associated with either the maximum or minimum eigenvalue is selected and applied to the electrode array as the subsequent DEP actuation pattern. The resulting drive signals generate a non-uniform electric field that produces an optimal DEP force, thereby moving the particle in the chosen direction.}
\label{ED1}
\end{figure}

\begin{figure}[h]
\centering
\includegraphics[width=1\textwidth]{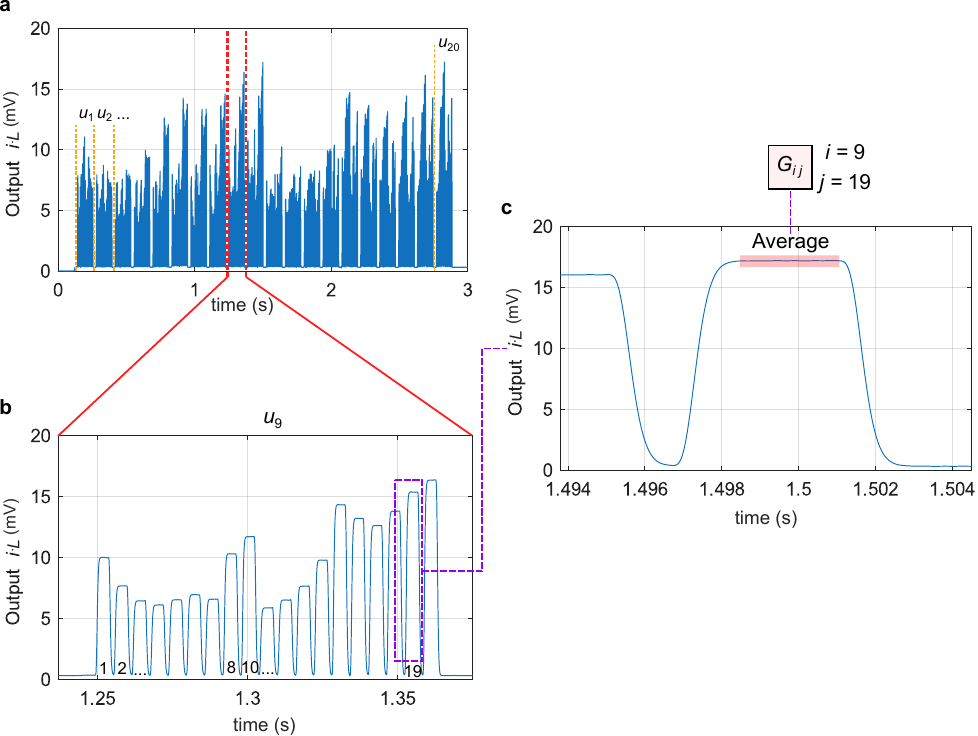}
\caption{\textbf{Electrode signals during conductance matrix acquisition.} (a) Full time trace of a complete conductance matrix acquisition, comprising 380 current measurements obtained via TIA with a gain $L = 100$~k$\Omega$ by sequentially applying 0.35~V peak-to-peak at 100~kHz excitation to each of the 20 electrodes and measuring the resulting currents from the remaining 19 electrodes. The diagonal elements were not measured in this implementation and were set to zero in the final matrix. (b) Signals recorded from all electrodes when electrode 9 is excited. (c) Zoomed-in view of the current corresponding to the coupling between electrodes 9 and 19. The associated conductance element $G_{9,19}$ is obtained by averaging the steady-state plateau value of the signal and normalizing it by the applied input voltage. }
\label{ED2}
\end{figure}

\begin{figure}[h]
\centering
\includegraphics[width=1\textwidth]{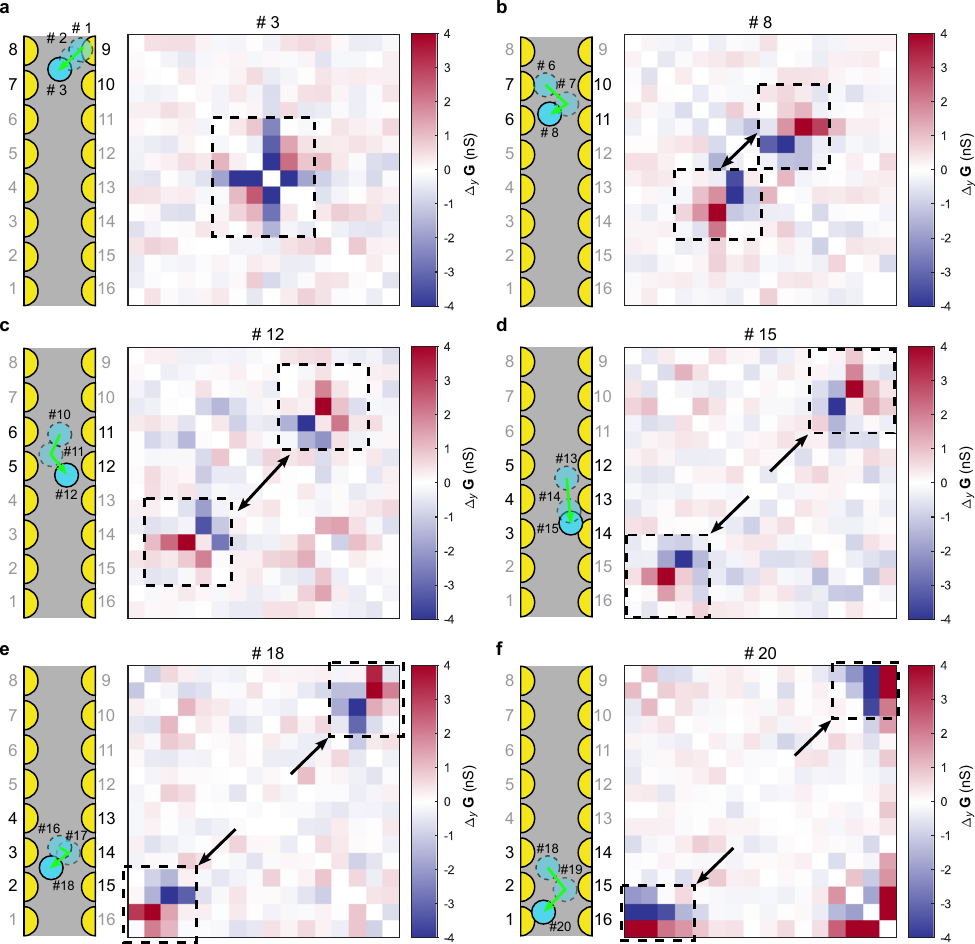}
\caption{\textbf{Conductance matrix evolution during particle motion}. Particle motion during the experiment presented in Figure~\ref{f2} is further studied to retrieve particle position relative to the electrodes and the corresponding changes in the the conductance matrices across iterations \#1--3 (a,~b), \#6--8 (c,~d), \#10--12 (e,~f), \#13--15 (g,~h), \#16--18 (i,~j), and \#18--20 (k,~l). }
\label{ED3}
\end{figure}

\begin{figure}[h]
\centering
\includegraphics[width=1\textwidth]{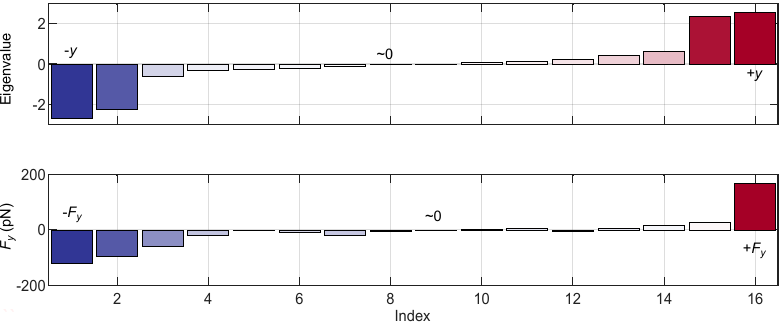}
\caption{\textbf{Eigenvalue-force relationship from the VC operator.} Simulation results showing the eigenvalues and corresponding force magnitudes obtained from the VC operator constructed using the iteration triplet \#9--11 of the experiment in Fig.~\ref{f2}b. The forces were simulated in COMSOL using the Minkowski stress tensor, with the corresponding eigenvectors applied to the electrodes as excitation voltages.  }
\label{ED4}
\end{figure}

\begin{figure}[h]
\centering
\includegraphics[width=1\textwidth]{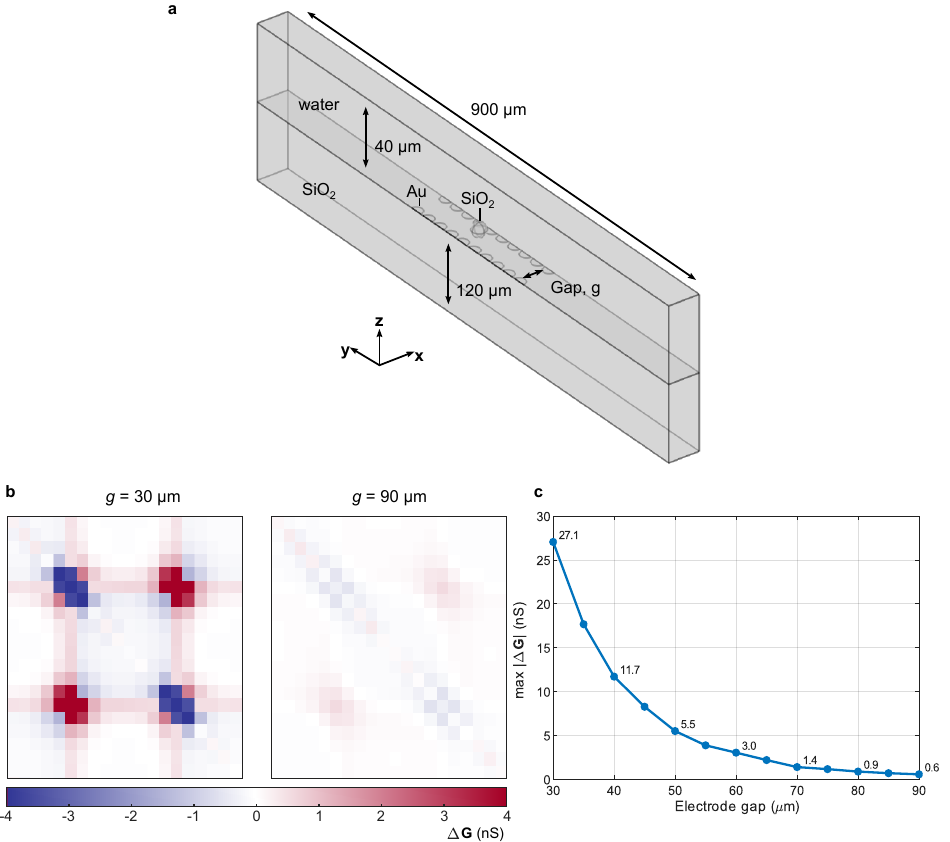}
\caption{\textbf{Conductance matrix sensitivity on the electrode distance.} (a) Three-dimensional COMSOL geometry of a straight microfluidic channel used for one-dimensional manipulation. The channel (900$\times$120$\times$40 $\mu$m$^3$, water) is enclosed by PDMS walls, with coplanar Au electrodes patterned on a SiO$_2$ substrate and separated by a variable gap $g$. A dielectric SiO$_2$ particle is placed at the channel center between the electrodes. (b) The matrix $\Delta \mathbf{G}$ is defined as the difference between conductance matrices measured with ($\mathbf{G}_\text{particle}$) and without ($\mathbf{G}_\text{empty}$) the particle between the electrodes, $ \Delta \mathbf{G} = \mathbf{G}_\text{particle} - \mathbf{G}_\text{empty}$, for two representative electrode gaps, $g = 30$~$\mu$m (left) and $g = 90$~$\mu$m (right). For the smaller gap, the particle induces pronounced conductance perturbations across multiple electrode pairs, whereas at $g = 90$~$\mu$m the signal is attenuated by approximately one order of magnitude. (c) Maximum absolute conductance change, $\max |\Delta \mathbf{G}|$, as a function of electrode gap. The sensitivity decreases monotonically from 27.1~nS at $g = 30$~$\mu$m to 0.6~nS at $g = 90$~$\mu$m. }
\label{Avsg}
\end{figure}

\begin{figure}[h]
\centering
\includegraphics[width=1\textwidth]{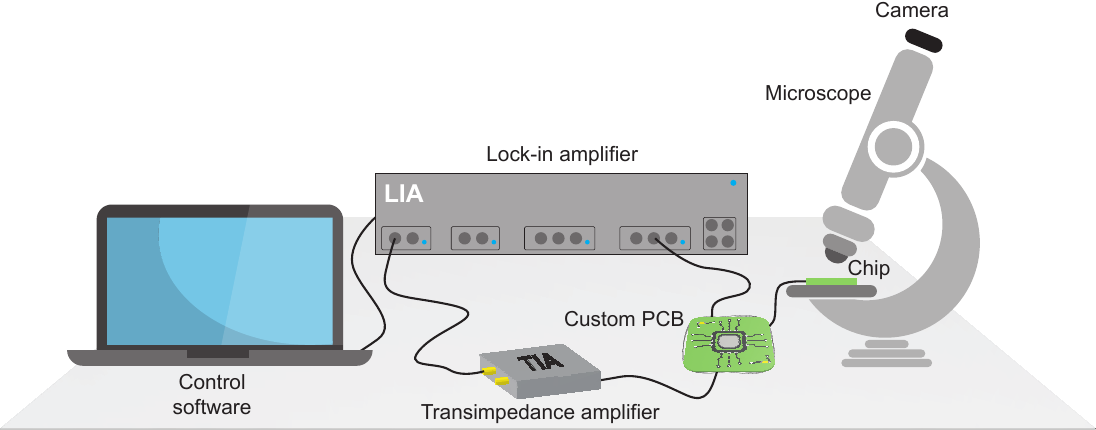}
\caption{\textbf{Schematic illustration of the experimental setup used for DGEM}. The system comprises a computer running the control software, a lock-in amplifier for signal generation and demodulation, a transimpedance amplifier for electrical current measurements, and a custom PCB integrating a microcontroller with circuitry for signal amplitude control and switching. The setup also includes a microfluidic chip with particles, and a microscope equipped with a digital camera for real-time particle tracking. }
\label{ED5}
\end{figure}

\begin{figure}[h]
\centering
\includegraphics[width=1\textwidth]{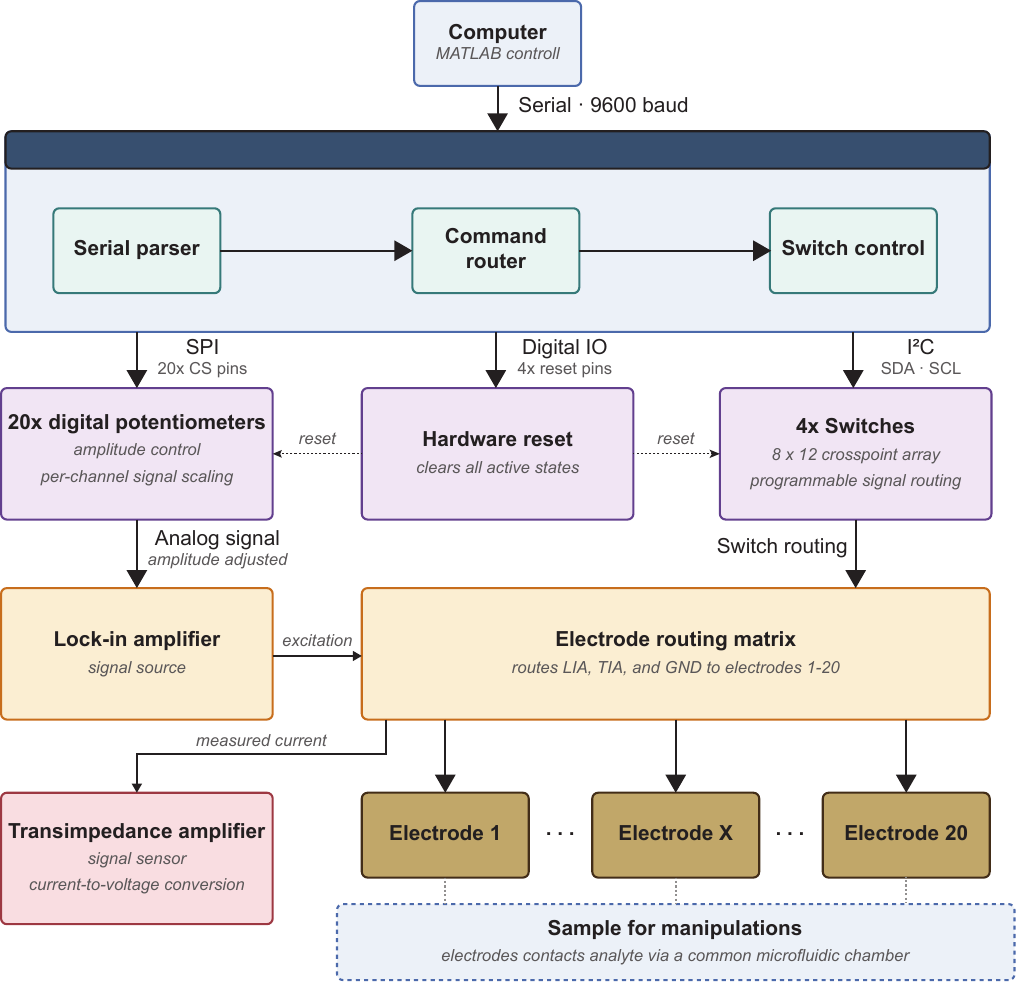}
\caption{\textbf{Functional block diagram of the experimental setup.} Experiment parameters and commands are generated by the master control software (MATLAB) running on the host computer and transmitted to the Arduino microcontroller over a 9600-baud serial link. The Arduino processes the incoming data through three main stages: (i) a serial parser that extracts integer values from the byte stream, typically corresponding to electrode driving amplitudes, (ii) a command router that selects the operation mode (e.g., electrode excitation or sensing), and (iii) a matrix-measurement control layer that configure analog switches to ensure correct electrode addressing during response measurements. The Arduino interfaces with external hardware through SPI, I$^2$C, and digital I/O lines. The SPI bus controls the wiper positions of 20 digital potentiometers, which set the per-channel excitation amplitude supplied by the lock-in amplifier (LIA). Four digital I/O lines are used to reset the wiper positions and/or return the analog switches to a known state. An I$^2$C bus configures four analog crosspoint switches. Together, these elements enable routing of the LIA, TIA, and ground connections to any of the 20 electrodes in contact with the sample. Finally, the TIA returns the current measured at the selected electrode to the LIA for demodulation of the measured current signal. }
\label{ED6}
\end{figure}

\begin{figure}[h]
\centering
\includegraphics[width=1\textwidth]{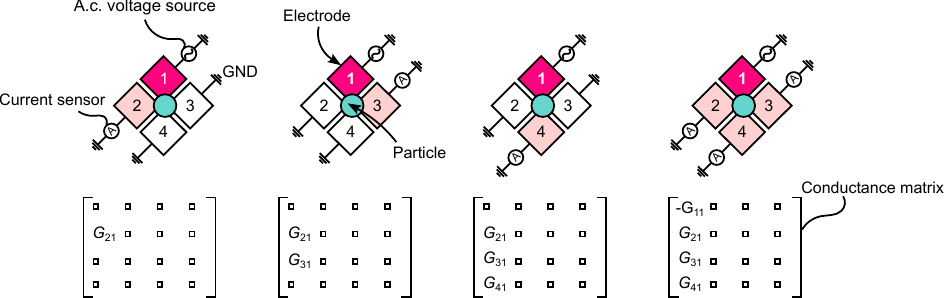}
\caption{\textbf{Schematic illustration of the conductance matrix acquisition process.} The electrodes are sequentially connected to the a.c.\ voltage source (0.35~V peak-to-peak at 100~kHz), a current sensor (transimpedance amplifier with a gain of 100 k$\Omega$), or ground (GND). To measure the off-diagonal matrix element, one electrode in the array is driven by the excitation signal while all other electrodes are grounded, except one, which is connected to the sensing circuit. This configuration enables measurement of the conductance between the driven electrode and the selected sensing electrode. Diagonal conductance elements are measured by exciting a single electrode while connecting all remaining electrodes to the current-sensing circuit, thereby recording the aggregate electrical response. }
\label{ED7}
\end{figure}

\clearpage

\includepdf[pages=-]{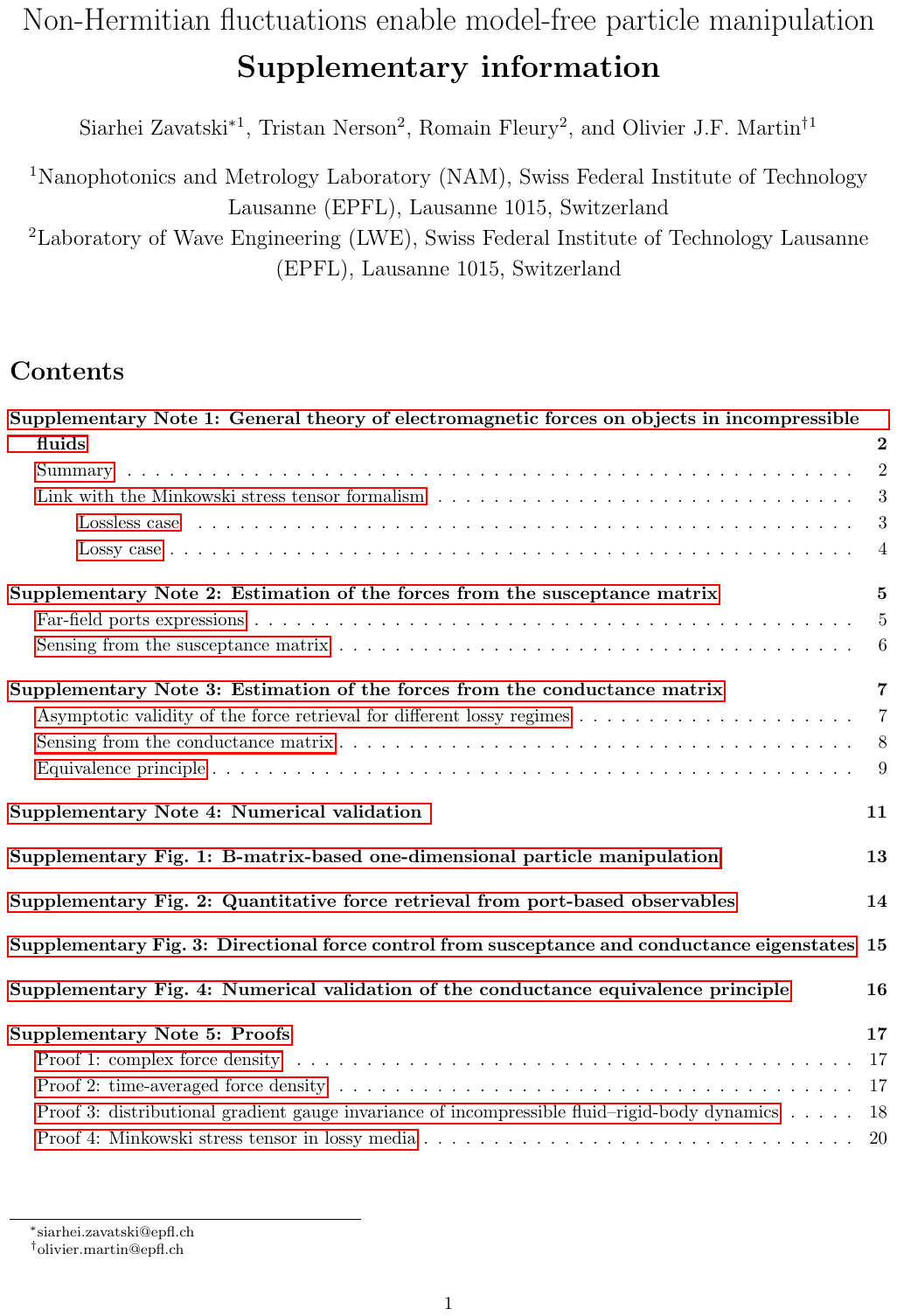}

\end{document}


\maketitle
\vspace{-0.5cm}
\tableofcontents

\newpage

\section{Supplementary Note 1: General theory of electromagnetic forces on objects in incompressible fluids}

\subsection{Summary}

Here, we introduce a self-consistent theory for electromagnetic forces in inhomogeneous, lossy, isotropic, linear media, that serves as the cornerstone of the experimental framework developed in this work. We follow the approach of Ref.\ \cite{nieto2022complex} (so far formulated only in vacuum), in which a complex electromagnetic force was defined through analytically continued fields. Using $\tau=t+js$ with $j^2=-1$, let $\mathbf E(\mathbf r,\tau), \mathbf H(\mathbf r,\tau), \mathbf B(\mathbf r,\tau), \mathbf P(\mathbf r,\tau)$ and $\mathbf M(\mathbf r,\tau)$ be analytic-signal fields satisfying Maxwell's equations \cite{kaiser2014completing}. We assume a linear isotropic medium with possibly spatially varying, perhaps complex susceptibilities, such that $\mathbf P=\varepsilon_0\chi_e\,\mathbf E, \mathbf M=\chi_m\,\mathbf H, \mathbf B=\mu\,\mathbf H, \varepsilon=\varepsilon_0(1+\chi_e)$ and $\mu=\mu_0(1+\chi_m)$. The free current obeys a local Ohm's law $\mathbf{J}_f=\sigma\mathbf{E}$, with spatially varying conductivity $\sigma \in \mathbb R$. For simplicity, the time-domain constitutive relations written here are understood to apply to dispersionless media; although temporal dispersion could in principle be incorporated through causal convolution kernels\cite{yuffa2012linear}, this would complicate some of the time-domain expressions, while being irrelevant for the present work since we restrict our attention to monochromatic regimes, including the experimental conditions considered here.

To extend the result of Ref.\ \cite{nieto2022complex} to lossy, inhomogeneous media, we begin from the analytic-signal counterpart of an expression in Ref. \cite{griffiths2015s} for the total force density, derived from the Lorentz force law:
\begin{equation}
\mathbf f^{\mathbb C}
=
\underbrace{\rho_f \mathbf E^*
+\mathbf J_f\times \mathbf B^*}_{\text{Free charges/currents}}
+\underbrace{(\mathbf P\cdot\nabla)\mathbf E^*
+\partial_\tau \mathbf P\times \mathbf B^*
+(\mathbf M\cdot\nabla)\mathbf B^*
+\mathbf M\times(\nabla\times\mathbf B^*)}_\text{Bound charges polarization and magnetization}.
\label{eq:analyticGriffiths}
\end{equation}
We adopt here the convention that, in the complex force density, the field factor is conjugated ($^*$) but the source factor is not. In the real-field limit, $s=0$, $\mu^*=\mu$, $\mathbf{E}^*=\mathbf{E}$, $\mathbf{B}^*=\mathbf{B}$, and it was shown by Anghinoni et al.\ \cite{anghinoni2023microscopic} that \cref{eq:analyticGriffiths} reduces to
\begin{equation}
    \mathbf f
    =
    \rho_f\mathbf E+\mathbf J_f\times\mathbf B
    +\frac12\nabla(\mathbf P\cdot\mathbf E)
    +\frac12\nabla(\mathbf M\cdot\mathbf B)
    -\frac12|\mathbf E|^2\nabla\varepsilon
    -\frac12|\mathbf H|^2\nabla\mu
    +\partial_t(\mathbf P\times\mathbf B)
\end{equation}
in lossless media. In lossy media, a similar derivation (see \textbf{Supplementary Note 5: Proof 1}) leads to the extended complex force density,
\begin{equation}
\begin{aligned}
\mathbf f^{\mathbb C}
={}&\rho_f \mathbf E^*
+\mathbf J_f\times \mathbf B^*
+\frac12\nabla(\mathbf P\cdot\mathbf E^*)
-\frac12|\mathbf E|^2\nabla\varepsilon
-j\varepsilon_0\chi_e\Im\mathcal Q_E 
\\
&\quad
+\chi_m|\mathbf H|^2\nabla\mu^*
+\frac{\mu^*\chi_m}{2}\nabla|\mathbf H|^2
-j\mu^*\chi_m\Im\mathcal Q_H +\partial_t(\mathbf P\times\mathbf B^*),
\end{aligned}
\label{eq:complex_force}
\end{equation}
with
\begin{equation}
    \mathcal Q_E := (\mathbf E^*\cdot\nabla)\mathbf E +\mathbf E^*\times(\nabla\times\mathbf E), \qquad \mathcal Q_H := (\mathbf H^*\cdot\nabla)\mathbf H +\mathbf H^*\times(\nabla\times\mathbf H).
\end{equation}

In the frequency domain, for a monochromatic analytic signal with angular frequency $\omega$, we write $\mathbf E(\mathbf r,\tau)=\widetilde{\mathbf E}(\mathbf r)e^{+j\omega\tau}$, and similarly for all other fields and sources. We show in \textbf{Supplementary Note 5: Proof 2} that the monochromatic time-averaged force density $\langle \mathbf f\rangle=\Re \widetilde{\mathbf f}^{\mathbb C} / 2$ admits the decomposition
\begin{equation}
\langle \mathbf f\rangle
= \mathbf f_{\mathrm{ng}} + \nabla \Pi,
\label{eq:time_averaged_total}
\end{equation}
with the pure-gradient part
\begin{equation}
\Pi
=
\frac14\Re\big(
\widetilde{\mathbf P}\cdot\widetilde{\mathbf E}^{*}
+
\widetilde{\mathbf M}\cdot\widetilde{\mathbf B}^{*}
\big),
\end{equation}
and the non-gradient part
\begin{equation}\boxed{
\begin{aligned}
\mathbf f_{\mathrm{ng}}
=&
\frac12\Re\left[
\frac{j\varepsilon}{1-j\theta}
(\widetilde{\mathbf E}\cdot\nabla\theta)
\widetilde{\mathbf E}^{*}
\right]
+\frac12\Re\left[
\sigma\widetilde{\mathbf E}\times\widetilde{\mathbf B}^{*}
\right]
\\
&\quad
-\frac14|\widetilde{\mathbf E}|^2\nabla\Re\varepsilon
+\frac12\Im\varepsilon\Im\widetilde{\mathcal Q}_E-\frac14|\widetilde{\mathbf H}|^2\nabla\Re\mu
+\frac12\Im\mu\Im\widetilde{\mathcal Q}_H
\end{aligned}} \quad ,
\label{eq:f_ng}
\end{equation}
where we have introduced the complex conduction-to-displacement ratio $\boxed{\theta=\sigma/(\omega \varepsilon)}$.

In \textbf{Supplementary Note 5: Proof 3}, we show that only the non-gradient force density $\mathbf{f}_\text{ng}$ contributes to the rigid-body dynamics of objects immersed in incompressible fluids. This implies that the total time-averaged net force acting on an object can be obtained solely from the volume integral of \cref{eq:f_ng}. In fact, the purely gradient contributions -- associated with electrostriction and magnetostriction force densities \cite{anghinoni2023microscopic} -- are known to vanish when evaluating the \emph{total} volumetric force rather than the local force density within matter \cite{brevik1979experiments,melcher1981continuum}, as they give rise only to  local deformations on deformable media. Our formulation is fully consistent with this viewpoint and, importantly, remains valid in lossy systems where standard energy-based approaches may fail. 

\subsection{Link with the Minkowski stress tensor formalism}
Since a vast body of literature approaches radiation forces from the view point of tensorial stress surface integrals, instead of density volume integrals, we provide here an explicit discussion on the connection between these two points of view, including in inhomogeneous lossy scenarios. 

\subsubsection{Lossless case}

\cref{eq:f_ng} relies on polarization and magnetization to introduce the standard terms involving $|\mathbf E|^2\nabla \Re\varepsilon$ and $|\mathbf H|^2\nabla \Re\mu$. In Ref.\ \cite{griffiths2015s}, it is shown that defining bound charges and currents yields the same total net force on objects, differing only by a pure-gradient term. In lossless dielectric media, where $\varepsilon$ and $\mu$ are real and $\sigma=0$, these are the only surviving contributions. We next discuss the ambiguous interpretation of their physical origin in the literature.

In Ref.\ \cite{griffiths2012resource}, Eq.~(59) is presented as the force density on the free charges and currents in the material,
\begin{equation}\label{eq:identity_s8}
\mathbf f_f=\rho_f\mathbf E+\mathbf J_f\times\mathbf B
=\nabla\cdot\mathbf T_M-\partial_t(\mathbf g_M),
\end{equation}
with the Minkowski stress tensor $\mathbf T_M$ given by
\begin{equation}
    \mathbf T_M =
\varepsilon \mathbf E\otimes\mathbf E
+
\mu \mathbf H\otimes\mathbf H
-
\frac{1}{2}
\left(
\varepsilon |\mathbf E|^2
+
\mu |\mathbf H|^2
\right)\mathbf I,
\end{equation}
with $\mathbf I$ the identity tensor, and $\mathbf g_M=\mathbf D\times\mathbf B$. Milonni and Boyd\cite{milonni2010momentum} follow essentially the same route: starting from the Lorentz force density on free charges and currents, they derive their Eqs.~(12)--(17), and later set the free sources to zero while retaining a nonzero force density in Eq.~(26). It appears that in a first step, the force on free charges and currents has been computed, in absence of a bound charge contribution. And then, the free charges and currents are set to zero, while finding  a nonzero result related to gradient of effective parameters, that are physically related to bound charges! To try to understand this paradox, one must recognize that in an inhomogeneous medium, \cref{eq:identity_s8} no longer holds as a purely free-source Lorentz force law. In isotropic linear lossless media one finds instead
\begin{equation}
\nabla\cdot\mathbf T_M-\partial_t(\mathbf g_M) = 
\rho_f\mathbf E+\mathbf J_f\times\mathbf B
-\frac12|\mathbf E|^2\nabla\varepsilon
-\frac12|\mathbf H|^2\nabla\mu.
\label{eq:brevik_correction}
\end{equation}
Brevik\cite{brevik1979experiments} states this explicitly in Eq.~(1.2a), where the first two terms on the right-hand side are the force on free charges and currents and the remaining terms act on the medium.

Confusion about the origin of the last two terms in \cref{eq:brevik_correction} has had consequences, for example in statements such as that of Ref.\ \cite{anghinoni2023microscopic} that ``Minkowski's formulation totally neglects the bound charges inside the material'', which does not hold in the sense explicitly mentioned above. The resulting picture is the following: the Lorentz force is the starting point; the Minkowski tensor appears as a useful macroscopic momentum-balance object derived from that starting point; and in lossless media it correctly reproduces $\mathbf f_\text{ng}$, provided one keeps track of the quantity that is actually being computed. After proper care of some potential discontinuities in $\mathbf T_M$ (see for example Ref.\ \cite{ye2017deriving}), we can check that the Minkowski surface integral is justified when the integration surface is chosen in a homogeneous exterior region. The time-averaged volume integral of force densities in \cref{eq:brevik_correction} gives the same net force on rigid bodies. 

\subsubsection{Lossy case}

In the lossy case, it is confirmed in \textbf{Supplementary Note 5: Proof 4} that
\begin{equation}\label{eq:minkowski_to_prove}
    \mathbf{f}_\text{ng} = \frac{1}{2} \Re[\nabla \cdot \widetilde{\mathbf{T}}_M] \equiv \frac{1}{2} \nabla \cdot \Re[\varepsilon \widetilde{\mathbf{E}} \otimes \widetilde{\mathbf{E}}^*+\mu \widetilde{\mathbf{H}} \otimes \widetilde{\mathbf{H}}^*] - \frac{1}{4} \Re[\nabla(\varepsilon |\widetilde{\mathbf{E}}|^2+\mu |\widetilde{\mathbf{H}}|^2)], \quad \text{with} \quad (\nabla \cdot A)_j = \partial_i A_{ij},
\end{equation}
which demonstrates that the Minkowski stress tensor is still perfectly valid when $\varepsilon$ and $\mu$ are complex to account for dielectric and magnetic losses. Here the Ohmic contribution is kept explicitly through
$\widetilde{\mathbf J}_{f}=\sigma\widetilde{\mathbf E}$.
Accordingly, the complex permittivity $\varepsilon$ entering the stress tensor
must not include the conductive term $-j\sigma/\omega$ and only describe bound charges.

It is important to specify the adopted convention for the divergence of a second-rank tensor, since it is not universal and the choice of contracted index determines the corresponding ordering of dyadic products such as $\widetilde{\mathbf{D}}\otimes\widetilde{\mathbf{E}}$ or $\widetilde{\mathbf{E}}\otimes\widetilde{\mathbf{D}}$, which impacts the equality for complex constitutive parameters. Similarly, in lossy media, a naively symmetrized Abraham-like stress tensor is not generally equivalent to the Minkowski stress tensor: the symmetrization removes antisymmetric contributions proportional to $\Im\varepsilon$ and $\Im\mu$, which are required to recover the non-gradient force density in \cref{eq:f_ng}. This distinction is often hidden in standard DEP theory\footnote{A notable exception is Sauer's work \cite{sauer1983forces} -- subsequently extended in fully vectorial lossy scattering calculations \cite{liu2026electromagnetic} -- which introduces an ad hoc symmetrized stress tensor within the momentum balance. Such manipulations must be interpreted carefully for complex and spatially inhomogeneous $\varepsilon$ and $\mu$. The resulting expression agrees with the present formulation only when the omitted loss-dependent antisymmetric dyadic terms vanish, e.g., for negligible magnetic effects and locally linearly polarized electroquasistatic fields.}, where the stress tensor derivation is typically performed either in the absence of dielectric and magnetic losses \cite{wang1997general}, or for locally linearly polarized electroquasistatic fields, where the loss-dependent antisymmetric dyadic contributions vanish. In general lossy electromagnetic media, however, it is the unsymmetrized Minkowski form that preserves the contributions arising from these $\Im\varepsilon$- and $\Im\mu$-dependent contributions.

\newpage
\section{Supplementary Note 2: Estimation of the forces from the susceptance matrix}

\subsection{Far-field ports expressions}

Here, we bridge the gap between microscopic electromagnetic force theory and macroscopic port-level measurements by showing how forces on particles can be inferred directly from variations observed at the system's control ports. This extends the framework developed in Ref.\ \cite{nerson2025multi}, where the force exerted on an object was estimated from changes in the scattering matrix under the assumption of  lossless media and the absence of  free charges and currents. Using the same notation, we define the immittance voltage ($\mathbf{u}$) and current ($\mathbf{i}$) vectors at some physically controllable ports (e.g., electric cables), such that $\mathbf{i}=\mathbf{Y}\mathbf{u}$, where $\mathbf{Y}$ denotes the admittance matrix. In electromagnetic systems where  all ports share the same characteristic impedance, the admittance matrix relating the immittance vectors is proportional to the conventional admittance matrix relating the physical voltages and currents measured in the ports. We further define a boundary $\mathbf{A}$, oriented outward from the microfluidic cell of volume $V$, such that it intersects all ports connected to the system. By computing small variations in $\mathbf{u}$ and $\mathbf{i}$ at the ports, it was shown that \cite{nerson2025multi}
\begin{equation}
\mathbf u^\dagger \delta\mathbf i-\delta\mathbf u^\dagger \mathbf i =
-\int_{\partial V}
\big(\widetilde{\mathbf E}\times \delta\widetilde{\mathbf H}^*-\delta\widetilde{\mathbf E}\times \widetilde{\mathbf H}^*\big) \cdot  \,\text{d}\mathbf{A} = - \int_V \nabla \cdot \big(\widetilde{\mathbf E}\times \delta\widetilde{\mathbf H}^*-\delta\widetilde{\mathbf E}\times \widetilde{\mathbf H}^*\big)\,\text{d}V.
\end{equation}
Using now Maxwell's equations with the more general assumptions of the \textbf{Supplementary Note 1}, leads to
\begin{equation}\begin{aligned}
        \Im[\mathbf u^\dagger \delta\mathbf i+\mathbf i^\dagger \delta\mathbf u]&=\Im[\mathbf u^\dagger \delta\mathbf i-\delta\mathbf u^\dagger \mathbf i] \\ &=\Im\int_V \Big[\delta\sigma\,|\widetilde{\mathbf E}|^2
-j\omega\,\delta\varepsilon^*\,|\widetilde{\mathbf E}|^2
-j\omega\,\delta\mu\,|\widetilde{\mathbf H}|^2 +\big(\sigma-j\omega\varepsilon^*\big)
\big(\widetilde{\mathbf E}\cdot\delta\widetilde{\mathbf E}^*-\delta\widetilde{\mathbf E}\cdot\widetilde{\mathbf E}^*\big) \\ & \qquad \qquad +j\omega\mu
\big(\delta\widetilde{\mathbf H}^*\cdot\widetilde{\mathbf H}-\widetilde{\mathbf H}^*\cdot\delta\widetilde{\mathbf H}\big)
\Big]\,\text{d}V.\end{aligned}
\end{equation}
Moreover, for a small rigid virtual displacement $\delta\mathbf r$ of an object, the constitutive parameters variations are exactly
\begin{equation}
    \delta\varepsilon=(\delta\mathbf r\cdot\nabla_{\delta \mathbf{r}})\varepsilon=-(\delta\mathbf r\cdot\nabla)\varepsilon,
\qquad
\delta\mu=(\delta\mathbf r\cdot\nabla_{\delta \mathbf{r}})\mu=-(\delta\mathbf r\cdot\nabla)\mu,
\qquad
\delta\sigma=(\delta\mathbf r\cdot\nabla_{\delta \mathbf{r}})\sigma=-(\delta\mathbf r\cdot\nabla)\sigma,
\label{eq:total_constitutive_gradient}\end{equation}
with $\nabla_{\delta \mathbf{r}}=(\partial_{\delta x}, \partial_{\delta y}, \partial_{\delta z})$. In general, however, displacing only some objects while keeping the surrounding objects, walls, and sources fixed does not imply \begin{equation}\nabla_{\delta \mathbf{r}} \widetilde{\mathbf{E}} = - \nabla \widetilde{\mathbf{E}} \quad \text{and} \quad \nabla_{\delta \mathbf{r}} \widetilde{\mathbf{H}} = - \nabla \widetilde{\mathbf{H}}\label{eq:total_gradient_delta_cdt}.\end{equation} Nevertheless, we note that these equalities remain approximately valid in typical  dielectrophoretic regimes. Consider an isolated particle embedded in a sufficiently lossy medium and located sufficiently far  from other particles and boundaries so that external perturbations to its response are negligible. If both the incident field and the particle-induced response vary only weakly  under a virtual displacement, e.g., $|\delta \mathbf{r}| \ll L_\text{inc}$ with $L_\text{inc} \sim |\widetilde{\mathbf{E}}_\text{inc}|/ |\nabla\widetilde{\mathbf{E}}_\text{inc}|$, then the field induced by the particle may be regarded as approximately rigidly translated under the virtual displacement.
It follows that $\widetilde{\mathbf{E}}_\text{ind}(\mathbf{r}, \delta \mathbf{r}) \approx \widetilde{\mathbf{E}}_\text{ind} (\mathbf{r}- \delta \mathbf{r}, \mathbf{0})$. We therefore obtain $\nabla_{\delta \mathbf{r}} \widetilde{\mathbf{E}}_\text{ind}=\nabla_{\delta \mathbf{r}} \widetilde{\mathbf{E}} \approx - \nabla \widetilde{\mathbf{E}}_\text{ind}$. Replacing $ \nabla \widetilde{\mathbf{E}}_\text{ind}$ by the total-field gradient $ \nabla \widetilde{\mathbf{E}}$ requires the stronger condition that $\nabla \widetilde{\mathbf{E}}_\text{inc} \approx \mathbf{0}$ locally. This local approximation is compatible with standard dielectrophoresis settings: the incident field may remain  nearly constant over the virtual displacement, while still exhibiting  a finite gradient over the larger characteristic length scale that determines the dielectrophoretic force\cite{pethig2017book}. The same reasoning applies to the $\widetilde{\mathbf{H}}$ field.

Under the above approximations, using \cref{eq:total_constitutive_gradient,eq:total_gradient_delta_cdt} together with standard vector identities yields, after simplification,
\begin{equation}
\frac{1}{4\omega}
\Im[\mathbf u^\dagger \delta\mathbf i+\mathbf i^\dagger \delta\mathbf u]
=
\delta\mathbf r\cdot
\int_{V}
\mathbf g_{\rm port}\,\text{d}V,
\end{equation}
with
\begin{equation}\label{eq:g_port}\boxed{
\mathbf g_{\rm port}
\equiv
\frac14|\widetilde{\mathbf E}|^2\nabla\Re\varepsilon
+\frac14|\widetilde{\mathbf H}|^2\nabla\Re\mu
+\frac{\sigma-\omega\Im\varepsilon}{2\omega}\Im\widetilde{\mathcal Q}_E
-\frac12\Im\mu\Im\widetilde{\mathcal Q}_H} \quad .
\end{equation}
Interestingly, the approximation introduced in \cref{eq:total_gradient_delta_cdt} is not required  in strictly lossless media. However, it becomes essential in the presence of  dielectric, magnetic, or Ohmic losses, as  the electromagnetic field variation terms contribute then directly to the imaginary port response.

\subsection{Sensing from the susceptance matrix}

Here, we show that, under suitable conditions, the total force exerted on an object can be inferred from measurements of variations in the imaginary part of the admittance matrix $\mathbf{Y}$ (the susceptance matrix), induced by a small virtual displacement of the object. Although the lossless condition (A1), defined below, is not satisfied in the experimental  regimes investigated in this work, the derivation that follows remains a key theoretical step to understand the non-Hermitian regime. Specifically, it establishes the lossless reference result from which we derive, in \textbf{Supplementary Note~3}, the equivalence principle that ultimately justifies using conductance variations to retrieve forces in the lossy regimes relevant to our experiments.

We consider a system in which the voltage vector $\mathbf{u}$ is driving the ports. We then  compare two physical states that differ only by a virtual displacement of the object, while the voltage remains unchanged, i.e., $\delta \mathbf{u}=0$. Under these conditions,
\begin{equation}
    \Im[\mathbf u^\dagger \delta\mathbf i+\mathbf i^\dagger \delta\mathbf u] = \Im[\mathbf u^\dagger \delta\mathbf i] = \mathbf{u}^\dagger \Im[\delta \mathbf{Y}] \mathbf{u} = \mathbf{u}^\dagger \delta \mathbf{B} \,\mathbf{u},
\end{equation}
where $\mathbf{B}$ is defined here as $\mathbf{B}= (\mathbf{Y}-\mathbf{Y}^\dagger) / 2j$. For reciprocal systems, the $\mathbf{Y}$-matrix is symmetric, and $\mathbf{B}$ is simply the elementwise imaginary part of $\mathbf{Y}$, i.e., the usual susceptance matrix. The quadratic form $\mathbf{u}^\dagger \delta \mathbf{B} \,\mathbf{u}$ then describes the fixed-drive differential reactive power in the system.

We now assume the following conditions: 
\begin{enumerate}
\item[(A1)] $\sigma \to 0$ in both object and the medium, with $\nabla \sigma \to 0$,
\item[(A2)] $\Im\varepsilon=0$, outside of the volume $V_{\rm obj}$ of an object,
\item[(A3)] $\Im\mu=0$, outside of the volume $V_{\rm obj}$ of an object.
\end{enumerate}
Then
\begin{equation}\label{eq:g_port_fng}\boxed{
    \mathbf f_{\rm ng}=-\,\mathbf g_{\rm port}} \quad .
\end{equation}
Finally, the total non-gradient force exerted on the object $\mathbf{F}_\text{ng}$ in the direction of a virtual displacement is
\begin{equation}\label{eq:delta_B} \boxed{
\mathbf{F}_\text{ng} \cdot \delta \mathbf{r} =
    \delta\mathbf r\cdot\int_{V_{\rm obj}}\mathbf f_{\rm ng}\,\text{d}V
=
-\delta\mathbf r\cdot\int_{V_{\rm obj}}\mathbf g_{\rm port}\,\text{d}V
=
-\frac{1}{4\omega}\mathbf{u}^\dagger \delta \mathbf{B} \,\mathbf{u}} \quad .
\end{equation}
Hence, under the absence of losses outside $V_{\rm obj}$ (permitting to reduce the volume integrals from $V$ to $V_{\rm obj}$), the force exerted on the object for a fixed voltage applied to the ports, can be directly computed from the variation of the imaginary part of the admittance matrix when this object is slightly displaced. The formula tolerates dielectric (bound charge) losses in the object. Yet, this is often not needed in dielectrophoresis regimes, because the dielectric loss is negligible compared to the Ohmic loss; dielectric losses of water due to the dipolar relaxations occurs at frequencies of tens of GHz\cite{ellison2007permittivity}, far beyond the typical range used in dielectrophoresis. The assumption (A1), however, is more subtle. In the \textbf{Supplementary Note 3}, we study regimes where $\sigma \neq 0$. We show that in some cases, \cref{eq:g_port_fng} can still hold, although a robust force estimation can be obtained using the real part of the admittance, i.e., the conductance matrix, which is used experimentally.

\newpage

\section{Supplementary Note 3: Estimation of the forces from the conductance matrix}

\subsection{Asymptotic validity of the force retrieval for different lossy regimes}

The experimental conditions of our dielectrophoresis experiments allow us to neglect dielectric losses, which are much smaller than Ohmic losses, while we explicitly include Ohmic losses through $\sigma$. The $\theta \in \mathbb R$ ratio is used as the standard criterion for good dielectrics/conductors, i.e.,  $\theta \ll 1$ for good dielectric material, while $\theta \gg 1$ for good conductors. We also use the electroquasistatic approximation to neglect terms involving $\widetilde{\mathbf{B}}$, $\widetilde{\mathbf{H}}$ and the magnetic material response. The port density of \cref{eq:g_port} simplifies as
\begin{equation}
\mathbf g_{\rm port}
\approx
\frac14|\widetilde{\mathbf E}|^2\nabla\varepsilon
+\frac{\varepsilon \theta}{2}\Im[(\widetilde{\mathbf E}^*\cdot\nabla)\widetilde{\mathbf E}].
\end{equation}
Simultaneously, the non-gradient force density in \cref{eq:f_ng} becomes
\begin{equation} \label{eq:simplified_fng}
\mathbf f_{\mathrm{ng}}
= -\frac14|\widetilde{\mathbf E}|^2\nabla\varepsilon+
\frac12\Re \Bigg[ \underbrace{
\frac{j\varepsilon}{1-j\theta}
(\widetilde{\mathbf E}\cdot\nabla\theta)}_{\widetilde{\rho}_f}
\widetilde{\mathbf E}^{*}
\Bigg].
\end{equation}

At material interfaces, the second term in \cref{eq:simplified_fng} must be understood in a distributional sense. When $\theta$ is discontinuous, the bulk free charge reduces to an effective free surface charge density $\widetilde{\rho}_s$. The resulting interfacial traction is therefore determined by the electric field acting on this surface charge sheet. Because  the tangential component of the electric field is continuous across the interface whereas  the normal component is generally discontinuous, we adopt the standard symmetric thin-layer prescription,
\begin{equation}
\widetilde{\mathbf E}_{\rm avg}
=
\widetilde{\mathbf E}_t+\frac{E_{1n}+E_{2n}}{2}\hat{\mathbf n},
\qquad
\mathbf t_\theta
=
\frac12\Re\left(\widetilde{\rho}_s\widetilde{\mathbf E}_{\rm avg}^{*}\right).
\end{equation}
Let $\Sigma$ be an interface separating two piecewise homogeneous media, labeled $k=1,2$, with unit normal $\hat{\mathbf n}$ oriented from medium 1 to medium 2, and write $\widetilde{\mathbf E}_k=E_{kn}\hat{\mathbf n}+\widetilde{\mathbf E}_t$ on both sides of the interface. Integrating the monochromatic charge-conservation law $\nabla\cdot\left[(\sigma+j\omega\varepsilon)\widetilde{\mathbf E}\right]=0$ over an infinitesimal pillbox crossing $\Sigma$ indicates that the normal total-current flux is continuous. We therefore define its unique value as
\begin{equation}
K_n
=
\hat{\mathbf n}\cdot(\sigma_k+j\omega\varepsilon_k)\widetilde{\mathbf E}_k
=
j\omega\varepsilon_k(1-j\theta_k)E_{kn},
\qquad k=1,2,
\end{equation}
where the equality holds irrespective of the side $k$ from which it is evaluated. Hence,
\begin{equation}
E_{kn}
=
\frac{K_n}{j\omega\varepsilon_k(1-j\theta_k)},
\end{equation}
and the free surface charge density is
\begin{equation}
\widetilde{\rho}_s
=
\varepsilon_2E_{2n}-\varepsilon_1E_{1n}
=
\frac{K_n}{j\omega}
\left( C(\theta_2) - C(\theta_1)\right), \qquad \text{with} \quad C(\theta) \equiv \frac{1}{1-j\theta}.
\end{equation}
Thus, the strength of the interfacial $\theta$-gradient force is controlled by the contrast of the complex screening factor $C(\theta)$, through $\widetilde{\rho}_s$, and not directly by the contrast of $\theta$ itself.

In the good dielectric high-frequency limit, $\theta\ll1$,
\begin{equation}
C(\theta)\approx1+j\theta,
\end{equation}
and therefore
\begin{equation}
\widetilde{\rho}_s
\approx
\frac{K_n}{\omega}(\theta_2-\theta_1) \propto \frac{K_n}{\omega^2}.
\end{equation}
Hence, when the two materials on either side of the interface are good dielectrics, the surface charge induced by a conductivity contrast is small and scales as $1/\omega^2$. In this regime, the $\theta$-gradient contribution to $\mathbf f_{\rm ng}$ represents a higher-order correction, while the additional $\theta$-dependent term in $\mathbf g_{\rm port}$ is also small. Therefore, to leading order, one recovers
\begin{equation}\label{eq:good_dielectric}
\mathbf g_{\rm port}\approx-\mathbf f_{\rm ng}.
\end{equation}

On the other hand, in the opposite good conductor limit, $\theta\gg1$, $C(\theta)\approx j/\theta$, so that
\begin{equation}
\widetilde{\rho}_s
\approx
\frac{K_n}{\omega}
\left(
\frac{1}{\theta_2}
-
\frac{1}{\theta_1}
\right) = K_n \left(\frac{\varepsilon_2}{\sigma_2}-\frac{\varepsilon_1}{\sigma_1}\right).
\end{equation}
In this regime, for fixed $K_n$, the normal displacement $\varepsilon_k E_{kn}$ vanishes as $1/\theta_k$, since the total current is carried predominantly by the Ohmic term. Thus, when the two materials on either side of the interface are good conductors and the $C(\theta)$ contrast is small, the interfacial free-charge traction associated with the $\theta$-gradient term can be small. However, this is not sufficient by itself to guarantee $\mathbf g_{\rm port}\approx-\mathbf f_{\rm ng}$, because $\mathbf g_{\rm port}$ still contains the phase-gradient term proportional to $\Im[(\widetilde{\mathbf E}^*\cdot\nabla)\widetilde{\mathbf E}]$. However, this term vanishes when the electric field has no spatially varying complex phase. In the electroquasistatic regime, the field is irrotational and satisfies
\begin{equation}\label{eq:electrostatic}
\nabla\cdot\left[(\sigma+j\omega\varepsilon)\nabla\phi\right]=0,
    \qquad
    \widetilde{\mathbf E}=-\nabla\phi .
\end{equation}
A sufficient condition for the field to exhibit no spatially varying complex phase is that $\theta$ be approximately uniform in space (i.e., $\theta_1 \approx \theta_2$), and that all imposed electrode voltages remain in phase. Indeed, if $\theta(\mathbf r,\omega)\approx\theta_0(\omega)$, then
\begin{equation}
\sigma+j\omega\varepsilon
=
\omega\varepsilon\left(\theta_0+j\right).
\end{equation}
The complex factor $(\theta_0+j)$ is then spatially uniform and can be factored out of the electroquasistatic equation \eqref{eq:electrostatic}. The remaining equation is a real electrostatic problem. Consequently, when electrode voltages are all in phase, the electroquasistatic potential can be globally rephased to become entirely real, and similarly for the electric field itself, i.e., $\widetilde{\mathbf E}=e^{j\alpha}\mathbf E_0$ with $\mathbf E_0\in\mathbb R^3$. It follows immediately that
\begin{equation}
\Im[(\widetilde{\mathbf E}^*\cdot\nabla)\widetilde{\mathbf E}]
=
\Im[(\mathbf E_0\cdot\nabla)\mathbf E_0]
=0.
\end{equation}
Hence, under these additional conditions, the good conductor limit is compatible with $\mathbf g_{\rm port}\approx-\mathbf f_{\rm ng}$.

Finally, for a strongly asymmetric interface, for example $\theta_2\gg1$ and $\theta_1\ll1$, one has $C(\theta_2)\approx0$ and $C(\theta_1)\approx1$, hence
\begin{equation}
\widetilde{\rho}_s\approx-\frac{K_n}{j\omega}\approx-\varepsilon_1 E_{1n}.
\end{equation}
The highly conducting side 2 therefore suppresses its normal electric displacement field, and the interface carries the charge required to terminate the displacement field coming from the dielectric side. Conversely, if $\theta_1\gg1$ and $\theta_2\ll1$, then $\widetilde{\rho}_s\approx \varepsilon_2 E_{2n}$. In such strongly asymmetric regimes, the interfacial $\theta$-gradient force is generally not negligible and one should not expect $\mathbf g_{\rm port}\approx-\mathbf f_{\rm ng}$ to hold.

\subsection{Sensing from the conductance matrix}

We now show that in a lossy regime, the force information can be accessed from the conductance matrix. Under the same uniform-$\theta$ condition introduced previously,
\begin{equation}
    \sigma(\mathbf r)+j\omega\varepsilon(\mathbf r)
    =
    \left(1+\frac{j}{\theta_0(\omega)}\right)\sigma(\mathbf r).
\end{equation}
The port currents are therefore all multiplied by the same factor $(1+j/\theta_0)$ with respect to the real conductive reference problem defined by $\sigma(\mathbf r)$. Hence, for reciprocal systems, the admittance matrix satisfies
\begin{equation}
    \mathbf{Y} = (1+j/\theta_0) \mathbf{G},
    \qquad
    \mathbf{G} \equiv \frac{\mathbf{Y}+\mathbf{Y}^*}{2}
    =
    \frac{\mathbf{Y}+\mathbf{Y}^\dagger}{2}.
\end{equation}
If $\theta_0(\omega)$ is unchanged by a virtual displacement, then
\begin{equation}
    \delta \mathbf{Y} = (1+j/\theta_0)\delta \mathbf{G},
\end{equation}
and therefore $\delta \mathbf{G}=\theta_0(\omega)\,\delta\mathbf{B}$. Here, $\mathbf{G}$ denotes the conductance matrix of the system. Consequently, the force acting on an object can be determined from measurements of the fixed-drive differential active (dissipative) power within the system:
\begin{equation}\label{eq:delta_G}
    \mathbf{F}_\text{ng} \cdot \delta \mathbf{r}
    \approx
    -\frac{1}{4\omega \theta_0(\omega)}
    \mathbf{u}^\dagger \delta \mathbf{G}\,\mathbf{u}.
\end{equation}
More importantly, mechanical information that is conventionally extracted from lossless, Hermitian scattering operators is here encoded in the non-Hermitian physics of the lossy admittance matrix: the force-related variations in susceptance can be inferred directly from dissipative conductance variations. Interestingly, Ohmic losses are not merely tolerated in this framework, but are \emph{essential} for such a measurements.

We now consider a final and particularly important regime: the low-frequency limit, in which $\sigma(\mathbf r)+j\omega\varepsilon(\mathbf r)\approx \sigma(\mathbf{r})$, and therefore $\mathbf{Y}\approx \mathbf{G}$. We show next that, even when $\theta(\mathbf{r})$ is not spatially uniform, the force can still be determined entirely from the conductance matrix.

\subsection{Equivalence principle}

The conductance-based force expression can be justified in the low-frequency regime through an argument that does not rely on the assumption of spatially uniform $\theta$. We denote by $\sigma(\mathbf r)$ the physical Ohmic conductivity and by $\varepsilon(\mathbf r)$ the physical permittivity. The electroquasistatic equation then reads 
\begin{equation}
\nabla\cdot\left[(\sigma+j\omega\varepsilon)\nabla\phi\right]=0,
\qquad
\widetilde{\mathbf E}=-\nabla\phi .
\end{equation}
In the low-frequency limit, assuming $\omega\varepsilon\ll\sigma$ in each phase, the potential $\phi$ satisfies, at leading order, the real conduction problem $\nabla\cdot\left(\sigma\nabla\phi\right)=0$. Let $\mathbf G$ be the conductance matrix associated with this limiting conduction problem. Thus, if $\mathbf u$ denotes the vector of imposed electrode voltages and $\boldsymbol{i}$ the corresponding vector of electrode currents in the conduction problem, we have $\boldsymbol{i}=\mathbf G\mathbf u$. In this limit,  no susceptance matrix appears, since the problem reduces to a purely real, zero-frequency conduction system.

We now introduce an auxiliary lossless dielectric problem. This auxiliary
problem has zero Ohmic conductivity ($\sigma^\sharp = 0$) and an artificial permittivity
\begin{equation}
\varepsilon^\sharp(\mathbf r)=\frac{\sigma(\mathbf r)}{\omega}.
\end{equation}
Its local admittance coefficient is therefore $j\omega\varepsilon^\sharp(\mathbf r) = j\sigma(\mathbf r)$. The auxiliary potential $\phi^\sharp$ satisfies $\nabla\cdot\left(j\omega\varepsilon^\sharp\nabla\phi^\sharp\right)=0$, or, using the definition of $\varepsilon^\sharp$ and dividing by $j$, $\nabla\cdot\left(\sigma\nabla\phi^\sharp\right)=0$. With the same electrode voltages, the auxiliary problem is therefore the same boundary-value problem as the limiting conduction problem. Hence $\phi^\sharp=\phi$ and $\widetilde{\mathbf E}^\sharp=\widetilde{\mathbf E}$. The auxiliary current density is
\begin{equation}
\widetilde{\mathbf J}^\sharp
=
j\omega\varepsilon^\sharp\widetilde{\mathbf E}^\sharp
=
j\sigma\widetilde{\mathbf E} = j \widetilde{\mathbf{J}},
\end{equation}
where $\widetilde{\mathbf{J}}$ is the current density in the limiting conduction problem. After integration over the ports, the same relation holds for the port currents:
\begin{equation}
\boldsymbol{i}^\sharp
=
j\boldsymbol{i}
=
j\mathbf G\mathbf u \equiv \mathbf{Y}^\sharp \mathbf u.
\end{equation}
Since this holds for any voltage vector $\mathbf u$, we obtain $\mathbf Y^\sharp=j\mathbf G \equiv j \mathbf B^\sharp$.

The auxiliary problem is lossless by construction, and therefore the susceptance-force relation in \cref{eq:delta_B} applies directly:
\begin{equation}
\mathbf F^\sharp\cdot\delta\mathbf r
=
-\frac{1}{4\omega}
\mathbf u^\dagger\delta\mathbf B^\sharp\mathbf u = -\frac{1}{4\omega}
\mathbf u^\dagger\delta\mathbf G\mathbf u .
\end{equation}
We now relate $\mathbf F^\sharp$ to the physical low-frequency force by evaluating the force on a closed surface entirely contained in the homogeneous exterior medium. In the physical problem, the exterior permittivity is $\varepsilon_m$, and the corresponding Minkowski stress tensor is given by
\begin{equation}
\langle \mathbf{T}_M \rangle
=
\frac{\varepsilon_m}{4}
\left(
2 \Re[\widetilde{\mathbf E}\otimes\widetilde{\mathbf E}^*]
-
|\widetilde{\mathbf E}|^2\mathbf I
\right).
\end{equation}
In the auxiliary problem, the exterior permittivity is
\begin{equation}
\varepsilon_m^\sharp=\frac{\sigma_m}{\omega},
\end{equation}
while the exterior field is the same. Therefore the auxiliary Maxwell stress is
\begin{equation}
\langle \mathbf{T}_M^\sharp \rangle
=
\frac{\varepsilon_m^\sharp}{4}
\left(
2\Re[\widetilde{\mathbf E}\otimes\widetilde{\mathbf E}^*]
-
|\widetilde{\mathbf E}|^2\mathbf I
\right).
\end{equation}
Thus
\begin{equation}
\langle \mathbf{T}_M \rangle
=
\frac{\varepsilon_m}{\varepsilon_m^\sharp}\langle \mathbf{T}_M^\sharp \rangle, \qquad  \mathbf F_{\rm ng}
=
\frac{\varepsilon_m}{\varepsilon_m^\sharp}\mathbf F^\sharp =
\frac{1}{\theta_m}\mathbf F^\sharp .
\end{equation}
Combining this result with the auxiliary susceptance relation yields the central equation of this Supplementary Information: \begin{equation}\label{eq:delta_G_bis}\boxed{
\mathbf F_{\rm ng}\cdot\delta\mathbf r
=
-\frac{1}{4\omega\theta_m}
\mathbf u^\dagger\delta\mathbf G\mathbf u} \quad .
\end{equation}
This derivation shows that the low-frequency validity of the conductance-based retrieval does not rely on the proportionality $\delta\mathbf G=\theta\delta\mathbf B$, which requires spatially uniform $\theta$. Instead, it follows from a distinct low-frequency correspondence: the physical field is governed by a real conduction problem, which can be formulated as an auxiliary lossless dielectric problem whose susceptance matrix coincides exactly with the limiting conductance matrix.

This also clarifies why the susceptance-based estimate need not achieve the same  accuracy  in the unmatched cases, even at low frequency. In this regime, $\mathbf G$ probes the response of the real conduction problem, which is governed by $\sigma$. By contrast, the physical susceptance matrix captures the first capacitive correction, which is governed by $\varepsilon$. These two responses are proportional only when $\varepsilon/\sigma$ is spatially uniform, equivalently when $\theta=\sigma/(\omega\varepsilon)$ is spatially uniform at fixed frequency. Consequently, when $\theta_p\neq\theta_m$, the conductance-based retrieval can remain accurate at low frequencies, whereas the susceptance-based retrieval generally exhibits a finite offset.

\newpage
\section{Supplementary Note 4: Numerical validation \label{simul}}

To test the manipulation procedure when either the susceptance matrix $\mathbf{B}$ or the conductance matrix $\mathbf{G}$ is used, we performed finite elements simulations using the AC/DC and fluid flow modules of COMSOL Multiphysics~6.4. The simulation domain geometry is depicted in Supplementary Fig.~\ref{s1}a. The spatial distribution of the electric field strength was simulated by solving the following equation, $-\nabla\cdot\left[(\sigma+j\omega\varepsilon)\nabla\phi\right]=0$, where $\phi$ is the electric potential at the angular frequency $\omega$, which relates to the electric field as $\widetilde{\mathbf{E}} = -\nabla \phi$. This field was further used to calculate the DEP force, $\mathbf{F}_\text{DEP}$, acting on the particle and to simulate its velocity, $\mathbf{v}$, and position, $\mathbf{r}$. The DEP force was calculated using the standard model implemented in COMSOL, while $\mathbf{v}$ and $\mathbf{r}$ were determined from the balance of DEP and drag forces, $\mathbf{F}_\text{drag}$:
\begin{equation}
    \frac{d}{dt} \left( m_p \frac{d\mathbf{r}}{dt}\right) = \mathbf{F}_\text{DEP}  +  \mathbf{F}_\text{drag}.
    \label{N2eq1}
\end{equation}

All simulations (except Fig.~\ref{s4}) were performed in a water background, assuming a real relative dielectric permittivity $\varepsilon_m = 78$ and an electrical conductivity $\sigma_m = 0.03$~S/m, which matches experimental conditions. Gold with $\sigma_\text{gold} = 45.6 \cdot 10^6$~S/m and $\varepsilon_\text{gold} = 6.9$ was used for the electrode material. A 20~$\mu$m--diameter silica particle was simulated with $\sigma_p = 10^{-4}$~S/m and $\varepsilon_p$ = 3.7. The simulations were performed in the frequency domain at 100~kHz, while the particle movement was simulated in the time domain, with a total simulation time of 2~ms and a time step of 0.02~ms. The minimum mesh size used in the geometry discretizations was 1~nm.

Supplementary Figs.~\ref{s1}b--f show the particle positions in the simulation domain at different iterations, using the $\mathbf{B}$-matrix. These images confirm that our method enables accurate manipulation of a 20 $\mu$m-diameter silica microbead when the force operator is constructed solely from the susceptance matrix. Simulations using the $\mathbf{G}$-matrix are given in the main text in Figs.~2(d-f).

Let us now compare the force predicted from the port-based formulations with the force obtained from the Minkowski stress tensor. Supplementary Fig.~\ref{s2} shows that both $\mathbf B$ and $\mathbf G$ contain information about the force, but that the conductance matrix $\mathbf G$ provides a more accurate estimate in lossy experimental regime under study. The data spread around the ideal relation (diagonal in Figs.~\ref{s2}b and c) is reduced for $\mathbf G$, compared to $\mathbf B$ and the distribution of relative errors is more tightly centered around zero; see Fig.~\ref{s2}d. Supplementary Fig.~\ref{s3} further confirms this trend in the resulting  force direction. When the actuation vector is selected using the operator constructed with $\mathbf G$, the resulting Minkowski force remains significantly more  aligned with the target $+y$ direction than when the operator is derived from $\mathbf B$. This behavior is consistent with the low-frequency conductance-based argument of \cref{eq:delta_G_bis}: in the present regime, the dominant electrical response is governed primarily by the conductive problem, making the conductance matrix a more reliable practical proxy for force optimization than the susceptance matrix.

Finally, Supplementary Fig.~\ref{s4} examines the frequency range over which the port-based force estimates reproduce the reference force obtained either from the non-gradient force density of Eq.~\ref{eq:simplified_fng} or from the Minkowski stress tensor. As demonstrated in  \textbf{Supplementary Note 1}, and confirmed numerically here, these two reference formulations -- the non-gradient force and the Minkowski force -- agree exactly across all simulations. The left column shows the susceptance-based estimate of \cref{eq:delta_B}, while the right column shows the conductance-based estimate of \cref{eq:delta_G_bis}. Under the matched-loss conditions $\theta_p=\theta_m$ [Figs.~\ref{s4}(a,b)], both port-based estimates remain consistent  with the reference forces across the entire frequency range. This behavior follows directly from the matched-loss condition discussed around \cref{eq:delta_G}: the admittance matrix is multiplied by a spatially uniform complex factor, so variations in $\mathbf B$ and $\mathbf G$ carry the same mechanical information. 

In this case, neither the susceptance nor the conductance formulation can be intrinsically privileged. When $\theta_p\neq\theta_m$, the two formulations become valid in different asymptotic regimes. For the susceptance matrix [Figs.~\ref{s4}(c,e)], agreement is recovered in the high frequency, corresponding to the good-dielectric limit $\theta\ll1$, where the conductivity-induced surface-charge contribution becomes a higher-order correction [\cref{eq:good_dielectric}]. The force is then dominated by the permittivity-gradient contribution, and the susceptance variation correctly captures the force. By contrast, for the conductance matrix [Figs.~\ref{s4}(d,f)], agreement is recovered at low frequency, corresponding to the good-conductor limit, $\theta \gg 1$. In this case, the equivalence principle of \cref{eq:delta_G_bis} maps this conductive problem to an auxiliary lossless dielectric problem whose susceptance matrix is exactly the limiting conductance matrix. Therefore, even when $\theta_p\neq\theta_m$, the low-frequency force can be retrieved from variations of the $\mathbf G$-matrix.

The key practical conclusion is therefore that, in the lossy low-frequency regime relevant to our experiments, $\mathbf G$ provides the more robust matrix. The susceptance matrix $\mathbf B$ remains effective in the matched-loss case or in the high-frequency dielectric limit, but it is not generally the optimal quantity when the leading-order field problem is conductive and the particle and surrounding medium possess  different $\theta$ values.

\newpage
\section[Supplementary Fig.~1: B-matrix-based one-dimensional particle manipulation]{Supplementary Fig.~1}
\begin{figure}[H]
      \centering
      \includegraphics[scale=1]{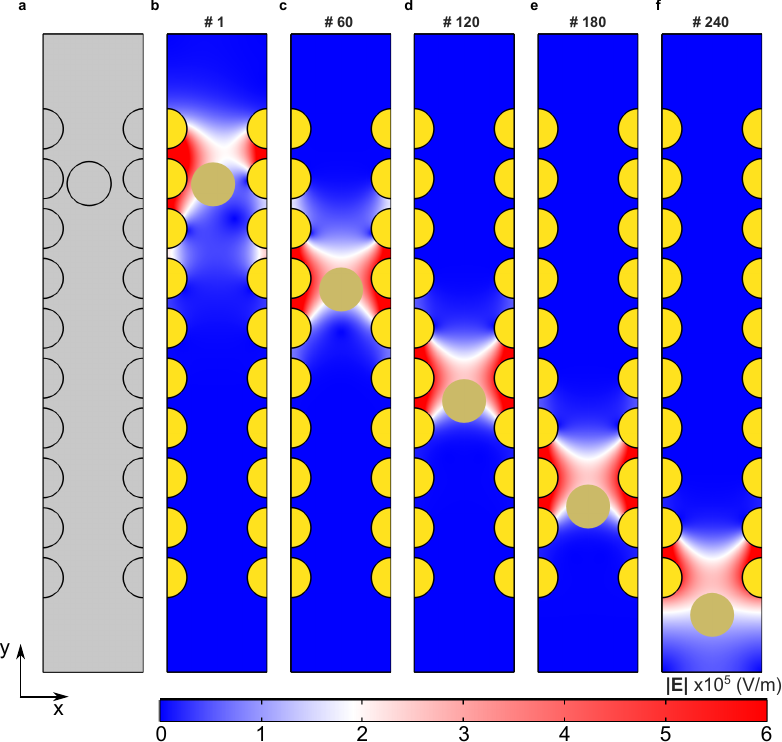}
      \caption{\textbf{B-matrix-based one-dimensional particle manipulation.} The variational susceptance operator of Eq.~\ref{eq:delta_B} is used to drive particle displacement. (a) Geometry of the simulation domain. Particle positions after iteration (b) \#~1, (c) \#~60, (d) \#~120, (e) \#~180, and (f) \#~240. The a.c.\ voltage has a maximum amplitude of 5~V peak-to-peak at the frequency of 100~kHz. The particle is displaced by 1~$\mu$m  between successive iterations.}
      \label{s1}
\end{figure}

\newpage
\section[Supplementary Fig.~2: Quantitative force retrieval from port-based observables]{Supplementary Fig.~2}

\begin{figure}[H]
      \centering
      \includegraphics[scale=0.9]{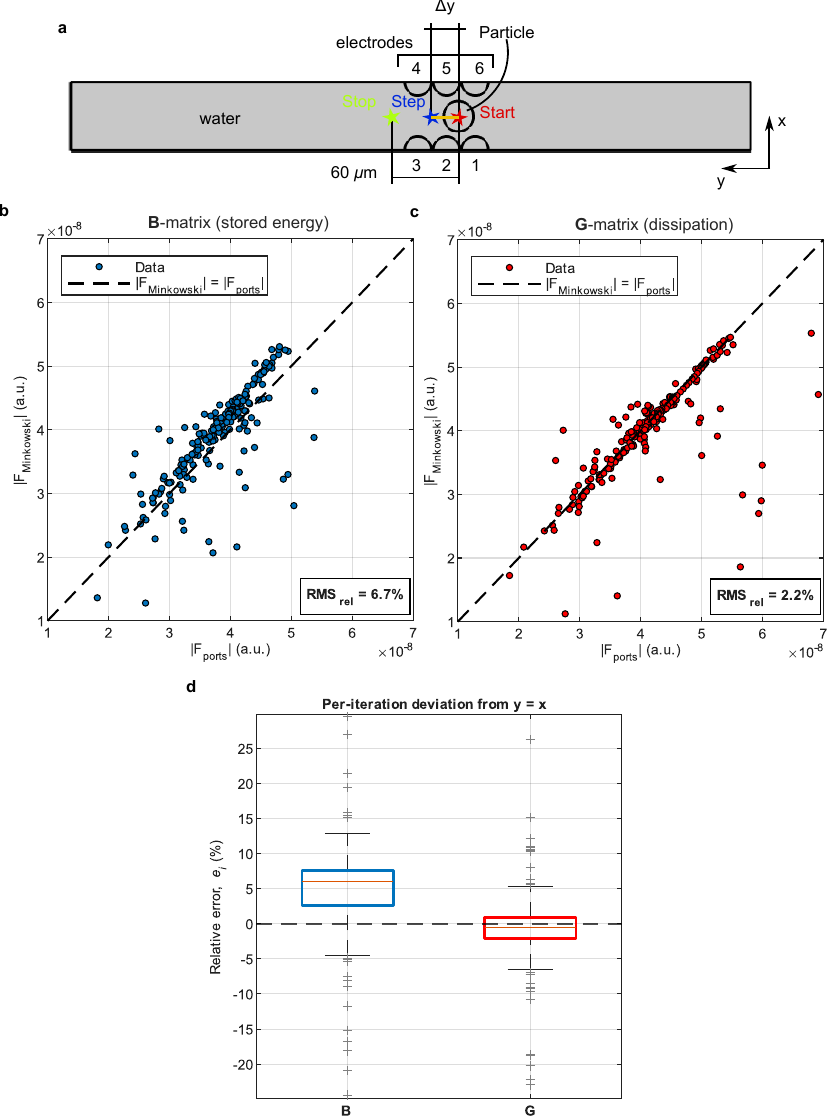}
      \caption{\textbf{Quantitative force retrieval from port-based observables.} Comparison of the $y$-component of the Minkowski stress tensor force, $F_\text{Minkowski}$, with the $y$-projection of the port-based force estimate $F_\text{port}$, calculated using the $\mathbf{B}$ or $\mathbf{G}$ matrix. (a) Schematic illustration of the geometry of the simulation domain. The particle was translated over a total distance of 60~$\mu$m with  steps of 0.25~$\mu$m. Scatter plots of $|F_\text{Minkowski}|$ versus $|F_\text{port}|$, when the latter was calculated (b) by Eq.~\ref{eq:delta_B} and (c) by Eq.~\ref{eq:delta_G_bis}, respectively; each point represents a single iteration (particle position). The dashed line indicates the ideal agreement ($|F_\text{Minkowski}| = |F_\text{port}|$). Inset statistics quantify the per-iteration root-mean-square relative error, defined as RMS$_\text{rel}=\sqrt{1/N\sum_{i=1}^Ne_i^2}$ with $e_i = (|F_{\text{Minkowski},i}| - |F_{\text{port},i}|)/|F_{\text{Minkowski},i}|$, where $N$ is the total number of iteration points lying within the 1.5$\times$interquartile range (IQR). (d) Box-and-whisker plot comparing the distributions of per-iteration relative error (in \%) between the $\mathbf{B}$- and $\mathbf{G}$-matrix cases. The central red line in the box marks the median, box edges span the IQR, showing data in 25th–75th percentiles, and whiskers extend to the most extreme data points within $1.5\times$IQR. Points beyond the whiskers are indicated in gray labels. The box outline colors correspond to the respective data markers in panels (a) and (b). }
      \label{s2}
\end{figure}

\newpage
\section[Supplementary Fig.~3: Directional force control from susceptance and conductance eigenstates]{Supplementary Fig.~3}

\begin{figure}[H]
      \centering
      \includegraphics[scale=0.9]{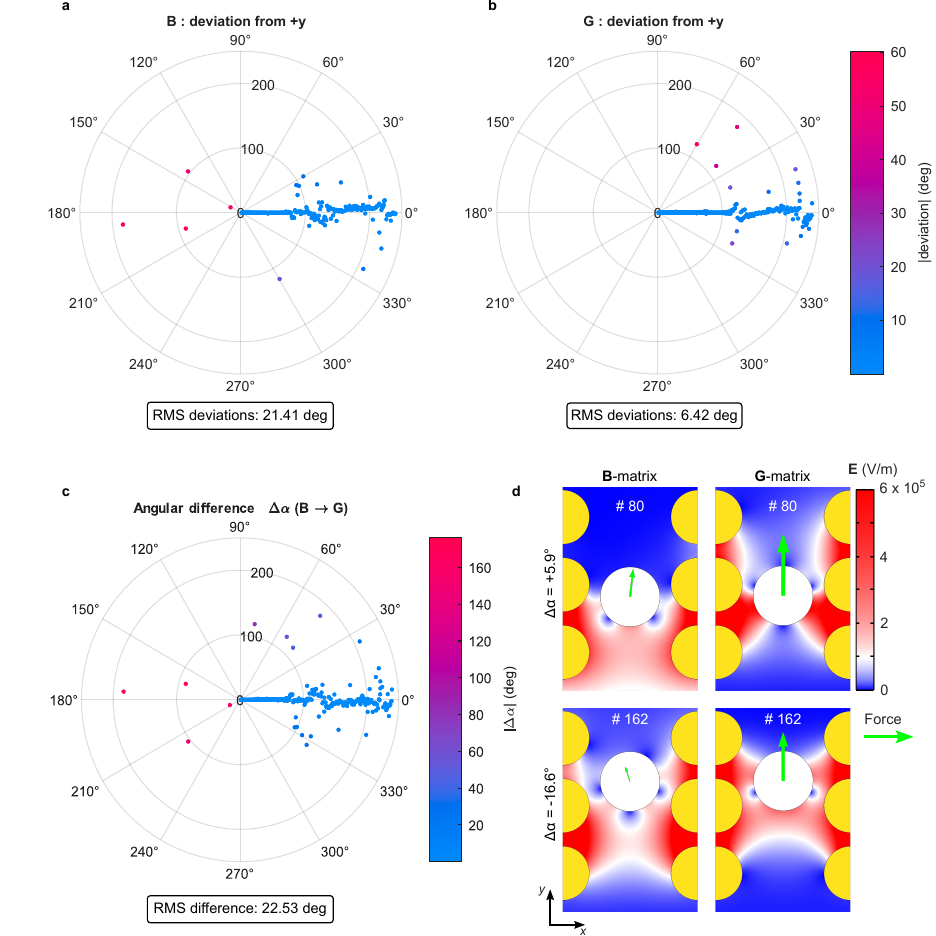}
      \caption{\textbf{Directional force control from susceptance and conductance eigenstates.} Directional analysis of the Minkowski stress tensor force vector, $\mathbf{F}_\text{Minkowski}$, calculated after applying the eigenvector corresponding to the maximum eigenvalue of the $\mathbf{R}_{\Delta y}$ operator, which is constructed from the $\mathbf{B}$- or $\mathbf{G}$-matrix. (a,~b) Polar scatter plots of the angular deviation of $\mathbf{F}_\text{Minkowski}$ from the positive $y$-axis when $\mathbf{R}_{\Delta y}$ uses (a) the $\mathbf{B}$-matrix or (b) the $\mathbf{G}$-matrix. Each point represents a single iteration (particle position). The angular coordinate represents the signed deviation angle, the radial coordinate corresponds to the iteration index, and the color encodes the absolute deviation magnitude. The RMS deviation is 21.41$\degree$ for $\mathbf{B}$- and 6.42$\degree$ for $\mathbf{G}$-matrix-based operators, indicating that the $\mathbf{G}$-matrix produces force vectors more consistently aligned with the vertical direction. (c) Polar scatter plot of the angular difference $\Delta \alpha$ between the $\mathbf{F}_\text{Minkowski}$ force vectors calculated using $\mathbf{B}$- and $\mathbf{G}$-based excitation signals. This difference is defined as the signed angle from the $\mathbf{B}$-vector to the $\mathbf{G}$-vector wrapped to $[-180\degree, 180\degree]$. The RMS difference of 22.53$\degree$ quantifies the overall directional discrepancy between both configurations. (d) COMSOL-simulated electric field magnitude distributions for two representative iterations (\#~80 and \#~162), shown for the $\mathbf{B}$-matrix (left column) and $\mathbf{G}$-matrix (right column). The green arrows indicate the direction of $\mathbf{F}_\text{Minkowski}$ acting on the SiO$_2$ particle (white circle). For iteration \#~80 the angular difference is $\Delta \alpha = +5.9\degree$, while for iteration \#~162, $\Delta \alpha = -16.6\degree$. The $\mathbf{G}$-matrix force vectors point nearly vertically in both cases, consistent with the tighter angular distribution observed in panel (b) compared to panel (a) for $\mathbf{B}$-matrix-based force vectors. }
      \label{s3}
\end{figure}

\newpage
\section[Supplementary Fig.~4: Numerical validation of the conductance equivalence principle]{Supplementary Fig.~4}

\begin{figure}[h!]
      \centering
      \includegraphics[scale=0.9]{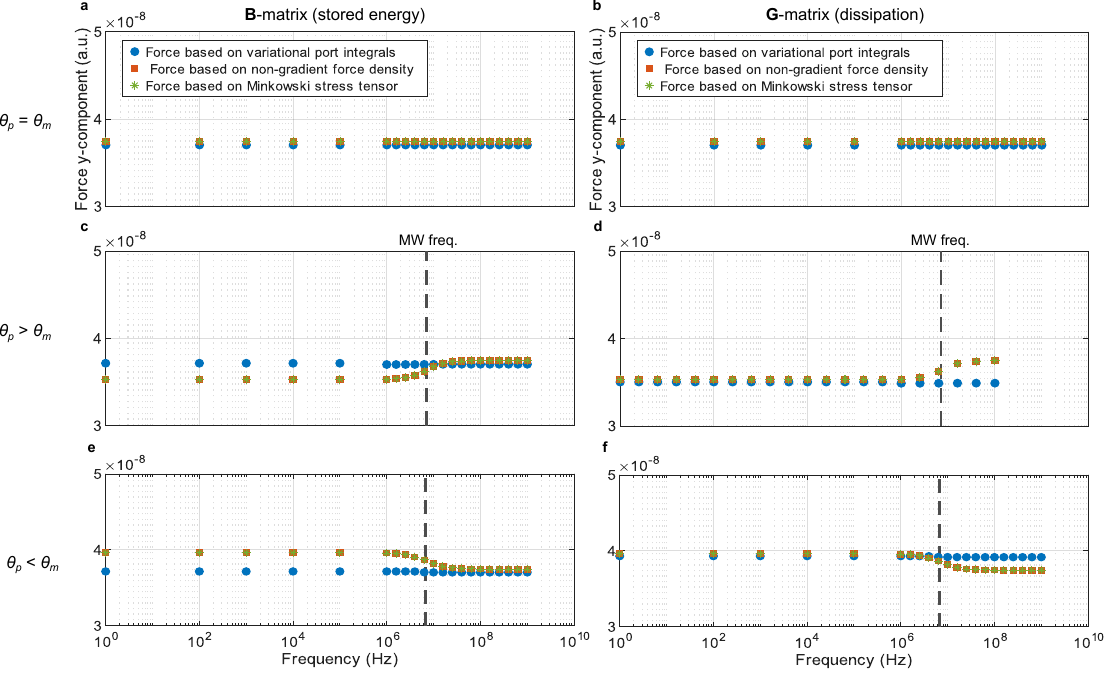}
      \caption{\textbf{Numerical validation of the conductance equivalence principle.} Frequency dependence of different force values: obtained by integrating Eq.~\ref{eq:simplified_fng} (orange squares), calculated by the variational port integrals (blue circles), and computed using the Minkowski stress tensor (green triangles), in different scenarios: (a,~b) when the ratio $\theta$ of particle and medium are equal ($\epsilon_p = 3.7$, $\sigma_p = 1.4\cdot10^{-3}$~[S/m] and $\epsilon_m = 78$, $\sigma_m = 0.03$~[S/m]), (c,~d) when $\theta_p > \theta_m$ ($\sigma_p = 2.8\cdot10^{-3}$~[S/m] and all other properties are the same as in a,~b),  and (e,~f) when $\theta_p < \theta_m$ ($\sigma_p = 1.4\cdot10^{-6}$~[S/m] and all other properties are the same as in a,~b). Eq.~\ref{eq:delta_B} was used to obtain the data in blue circles in (a,~c,~e), while Eq.~\ref{eq:delta_G_bis} was used to obtain the same data in (b,~d,~f). The geometry of the system used for simulation is the same as in Fig.~\ref{s2}a. The comparison shows that both formulations agree when $\theta_p=\theta_m$, while for unmatched $\theta$ ratios the $\mathbf{B}$-based estimate becomes accurate in the high-frequency dielectric regime and the $\mathbf{G}$-based estimate in the low-frequency conductive regime.}
      \label{s4}
\end{figure}

\newpage

\section{Supplementary Note 5: Proofs}

\subsection{Proof 1: complex force density}

Here, we prove \cref{eq:complex_force}. For the electric bound part, use
\begin{equation}
(\mathbf P\cdot\nabla)\mathbf E^*
=
\nabla(\mathbf P\cdot\mathbf E^*)
-(\mathbf E^*\cdot\nabla)\mathbf P
-\mathbf E^*\times(\nabla\times\mathbf P)
-\mathbf P\times(\nabla\times\mathbf E^*),
\end{equation}
together with $\mathbf P=\varepsilon_0\chi_e\mathbf E$ and the analytic Faraday law
\begin{equation}
\nabla\times\mathbf E=-\partial_\tau\mathbf B,
\qquad
\nabla\times\mathbf E^*=-\partial_{\tau^*}\mathbf B^*.
\end{equation}
After expanding the derivatives of $\chi_e$, the terms proportional to
$(\mathbf E^*\cdot\nabla\chi_e)\mathbf E$ cancel, and one obtains
\begin{equation}
(\mathbf P\cdot\nabla)\mathbf E^*
+\partial_\tau\mathbf P\times\mathbf B^*
=
\frac12\nabla(\mathbf P\cdot\mathbf E^*)
-\frac12|\mathbf E|^2\nabla\varepsilon
+\partial_t(\mathbf P\times\mathbf B^*)
-j\varepsilon_0\chi_e\Im\mathcal Q_E.
\end{equation}
For the magnetic bound part, use
\begin{equation}
(\mathbf M\cdot\nabla)\mathbf B^*
+\mathbf M\times(\nabla\times\mathbf B^*)
=
\nabla(\mathbf M\cdot\mathbf B^*)
-\mathbf B^*\times(\nabla\times\mathbf M)
-(\mathbf B^*\cdot\nabla)\mathbf M,
\end{equation}
together with $\mathbf M=\chi_m\mathbf H$ and $\mathbf B^*=\mu^*\mathbf H^*$. Expanding,
\begin{equation}
\nabla\times\mathbf M
=
\chi_m(\nabla\times\mathbf H)+(\nabla\chi_m)\times\mathbf H,
\end{equation}
so
\begin{equation}
\mathbf B^*\times(\nabla\times\mathbf M)
=
\mu^*\chi_m\mathbf H^*\times(\nabla\times\mathbf H)
+\mu^*|\mathbf H|^2\nabla\chi_m
-\mu^*(\mathbf H^*\cdot\nabla\chi_m)\mathbf H,
\end{equation}
and
\begin{equation}
(\mathbf B^*\cdot\nabla)\mathbf M
=
\mu^*(\mathbf H^*\cdot\nabla\chi_m)\mathbf H
+\mu^*\chi_m(\mathbf H^*\cdot\nabla)\mathbf H.
\end{equation}
The terms proportional to $(\mathbf H^*\cdot\nabla\chi_m)\mathbf H$ cancel, giving
\begin{equation}
(\mathbf M\cdot\nabla)\mathbf B^*
+\mathbf M\times(\nabla\times\mathbf B^*)
=
\nabla(\chi_m\mu^*|\mathbf H|^2)
-\mu^*|\mathbf H|^2\nabla\chi_m
-\mu^*\chi_m\mathcal Q_H.
\end{equation}
Since
\begin{equation}
\mathcal Q_H=\frac12\nabla|\mathbf H|^2+j\Im\mathcal Q_H,
\end{equation}
it follows that
\begin{equation}
(\mathbf M\cdot\nabla)\mathbf B^*
+\mathbf M\times(\nabla\times\mathbf B^*)
=
\chi_m|\mathbf H|^2\nabla\mu^*
+\frac{\mu^*\chi_m}{2}\nabla|\mathbf H|^2
-j\mu^*\chi_m\Im\mathcal Q_H.
\end{equation}
Adding the free-source terms proves the stated result.

\subsection{Proof 2: time-averaged force density}
Here, we prove \cref{eq:time_averaged_total}. We time-average \cref{eq:complex_force}. Since all terms involve the product of an analytic signal and a conjugated one, factors $e^{+j\omega(\tau-\tau^*)}=e^{-2\omega s}$ appear in all terms, which become independent of $t$. For such cross terms, $\partial_t=0$, and \cref{eq:complex_force} becomes
\begin{equation}
\widetilde{\mathbf f}^{\mathbb C}
=\widetilde{\rho}_f \widetilde{\mathbf E}^*
+\widetilde{\mathbf J}_f\times \widetilde{\mathbf B}^*
+\frac12\nabla(\widetilde{\mathbf P}\cdot\widetilde{\mathbf E}^*)
-\frac12|\widetilde{\mathbf E}|^2\nabla\varepsilon
-j\varepsilon_0\chi_e\Im\widetilde{\mathcal Q}_E +\chi_m|\widetilde{\mathbf H}|^2\nabla\mu^*
+\frac{\mu^*\chi_m}{2}\nabla|\widetilde{\mathbf H}|^2
-j\mu^*\chi_m\Im\widetilde{\mathcal Q}_H.
\end{equation}
Now, we can rewrite the free charge. Gauss' law gives $\widetilde{\rho}_f=\nabla\cdot(\varepsilon\widetilde{\mathbf E})$, while charge conservation reads
\begin{equation}
\nabla\cdot\widetilde{\mathbf J}_f + j\omega \widetilde{\rho}_f=\nabla\cdot(\sigma\widetilde{\mathbf E}) + j\omega \nabla\cdot(\varepsilon\widetilde{\mathbf E})=\nabla\cdot\big[(\sigma+j\omega\varepsilon)\widetilde{\mathbf E}\big]=0.
\end{equation}
Since $\sigma+j\omega\varepsilon=\omega\varepsilon(\theta+j)=j\omega\varepsilon(1-j\theta)$, we may write (for $\omega\neq0$)
\begin{equation}
\nabla\cdot\big[(1-j\theta)\varepsilon\widetilde{\mathbf E}\big]=0.
\end{equation}
Expanding the divergence yields
\begin{equation}
(1-j\theta)\nabla\cdot(\varepsilon\widetilde{\mathbf E})
+\varepsilon\widetilde{\mathbf E}\cdot\nabla(1-j\theta)=(1-j\theta)\widetilde{\rho}_f
-j\varepsilon\widetilde{\mathbf E}\cdot\nabla\theta=0,
\end{equation}
hence
\begin{equation}
\widetilde{\rho}_f
=
\frac{j\varepsilon}{1-j\theta}
(\widetilde{\mathbf E}\cdot\nabla\theta).
\label{eq:free_charge}
\end{equation}
Using Ohm's law, the total complex force density reads 
\begin{equation}
\begin{aligned}
\widetilde{\mathbf f}^{\mathbb C}
=&\frac{j\varepsilon}{1-j\theta}
(\widetilde{\mathbf E}\cdot\nabla\theta) \widetilde{\mathbf E}^*
+\sigma \widetilde{\mathbf E} \times \widetilde{\mathbf B}^*
+\frac12\nabla(\widetilde{\mathbf P}\cdot\widetilde{\mathbf E}^*)
-\frac12|\widetilde{\mathbf E}|^2\nabla\varepsilon
-j\varepsilon_0\chi_e\Im\widetilde{\mathcal Q}_E \\
&\quad +\chi_m|\widetilde{\mathbf H}|^2\nabla\mu^*
+\frac{\mu^*\chi_m}{2}\nabla|\widetilde{\mathbf H}|^2
-j\mu^*\chi_m\Im\widetilde{\mathcal Q}_H.   
\end{aligned}
\label{eq:full_expanded_complex_force}
\end{equation}
Now, we take half the real part of \cref{eq:full_expanded_complex_force}. Notably, we remark that
\begin{equation}
\frac14\Re\left[
\frac{|\widetilde{\mathbf H}|^2}{\mu_0}
\big((\mu-\mu_0)\nabla\mu^*-\mu^*\nabla\mu\big)
\right]
=
\frac14\frac{|\widetilde{\mathbf H}|^2}{\mu_0}
\Re\left[\big(2j\Im(\mu\nabla\mu^*)-\mu_0\nabla\mu\big)\right]
=
-\frac14|\widetilde{\mathbf H}|^2\nabla\Re\mu.
\end{equation}
After algebra, the time-averaged real force density finally becomes the stated \cref{eq:time_averaged_total}.

\subsection{Proof 3: distributional gradient gauge invariance of incompressible fluid--rigid-body dynamics}

Here, we show that the rigid-body dynamics of objects inside incompressible fluids are not affected by gradients of scalar fields in force densities expressions. We consider a fixed bounded domain $\Omega \subset \mathbb{R}^3$ occupied by an incompressible fluid region $\Omega_f(t)$ and finitely many rigid bodies $B_k(t)$, $k=1,\dots,N$, so that
\begin{equation}
\Omega=\Omega_f(t) \, \dot\cup \,\Big(\bigcup_{k=1}^N B_k(t)\Big).
\end{equation}
The velocity field $\mathbf v$ is continuous across fluid--solid interfaces, satisfies no-slip conditions, and is prescribed on $\partial\Omega$. In each rigid body,
\begin{equation}
\mathbf v(\mathbf x,t)=\mathbf U_k(t)+\boldsymbol\Omega_k(t)\times(\mathbf x-\mathbf X_k(t)),
\end{equation}
where $\mathbf x$ is the position, $\mathbf v$ the velocity field, $\mathbf U_k$ the translational velocity of body $k$, $\boldsymbol\Omega_k$ its angular velocity, and $\mathbf X_k$ its reference point (typically the center of mass).
The motion is incompressible, i.e., $\nabla\cdot\mathbf v=0$ in $\Omega$,
and in each phase the momentum balance is
\begin{equation}
\rho\mathbf a=\nabla\cdot\boldsymbol\sigma+\mathbf f,
\qquad
\mathbf a=\partial_t\mathbf v+\mathbf v\cdot\nabla\mathbf v,
\end{equation}
with stress decomposition
\begin{equation}
\boldsymbol\sigma=-p\mathbf I+\boldsymbol\tau.
\end{equation}

We now introduce a scalar field $\Pi(\mathbf x,t)$, assumed $C^1$ in each phase. On $\partial B_k(t)$, let $\Pi_f$ and $\Pi_k$ denote its traces on the fluid and rigid-body sides, respectively, and let $\mathbf n_k$ point out of $B_k(t)$ and into the fluid. We define
\begin{equation}
q_k:=\Pi_k-\Pi_f
\qquad\text{on }\partial B_k(t).
\end{equation}
We then transform the pressure and force density according to
\begin{equation}
p^\ast=p+\Pi,
\qquad
\mathbf f^\ast=\mathbf f+\nabla_{\mathcal D}\Pi,
\end{equation}
where $\nabla_{\mathcal D}\Pi$ is the global distributional gradient of the field $\Pi$, namely
\begin{equation}
\nabla_{\mathcal D}\Pi=(\nabla\Pi)_{\mathrm{bulk}}-\sum_{k=1}^N q_k\mathbf n_k\delta_{\partial B_k(t)},
\end{equation}
with
\begin{equation}
(\nabla\Pi)_{\mathrm{bulk}}=
\begin{cases}
\nabla\Pi_f, & \text{in }\Omega_f(t),\\
\nabla\Pi_k, & \text{in }B_k(t).
\end{cases}
\end{equation}
The key point is that this transformation changes only the isotropic part of the stress:
\begin{equation}
\boldsymbol\sigma^\ast=-p^\ast\mathbf I+\boldsymbol\tau
=\boldsymbol\sigma-\Pi\mathbf I.
\end{equation}
Since incompressibility does not involve $p$, changing $p$ does not affect whether $\mathbf{v}$ satisfies the kinematic constraints. In the bulk of each phase,
\begin{equation}
-\nabla p^\ast+\nabla\cdot\boldsymbol\tau+\mathbf f^\ast
=
-\nabla(p+\Pi)+\nabla\cdot\boldsymbol\tau+\mathbf f+\nabla\Pi
=
-\nabla p+\nabla\cdot\boldsymbol\tau+\mathbf f,
\end{equation}
so the classical bulk equations are exactly the same.

What remains is to check that the rigid-body force and torque are also unchanged. For body $B_k(t)$, the fluid traction is $\mathbf t_{f\to k}=\boldsymbol\sigma_f\mathbf n_k$. Under the transformation,
\begin{equation}
\boldsymbol\sigma_f^\ast=\boldsymbol\sigma_f-\Pi_f\mathbf I,
\end{equation}
so the traction changes by
\begin{equation}
\mathbf t_{f\to k}^\ast-\mathbf t_{f\to k}=-\Pi_f\mathbf n_k.
\end{equation}
Hence the change in interfacial force is
\begin{equation}
\Delta\mathbf F_{\partial B_k}
=
-\int_{\partial B_k(t)} \Pi_f\mathbf n_k\, \text{d}S.
\end{equation}
Inside the rigid body, the bulk part of the transformed force density contributes
\begin{equation}
\Delta\mathbf F_{B_k,\mathrm{bulk}}
=
\int_{B_k(t)} \nabla\Pi_k\, \text{d}V
=
\int_{\partial B_k(t)} \Pi_k\mathbf n_k\, \text{d}S,
\end{equation}
where we used the divergence theorem. Therefore the ordinary bulk-plus-traction contribution changes by
\begin{equation}
\Delta\mathbf F_{\mathrm{bulk/interface}}
=
\int_{\partial B_k(t)} (\Pi_k-\Pi_f)\mathbf n_k\, \text{d}S
=
\int_{\partial B_k(t)} q_k\mathbf n_k\, \text{d}S.
\end{equation}

But $\mathbf f^\ast$ was defined with the global distributional gradient, not only with the phasewise bulk gradients. Thus it also contains the singular interfacial term $-q_k\mathbf n_k\delta_{\partial B_k(t)}$, whose resultant is
\begin{equation}
\Delta\mathbf F_{\mathrm{sing}}
=
-\int_{\partial B_k(t)} q_k\mathbf n_k\, \text{d}S.
\end{equation}
These two contributions cancel exactly, so
\begin{equation}
\Delta\mathbf F_k=0.
\end{equation}

The torque calculation is identical. The ordinary traction and bulk terms produce
\begin{equation}
\Delta\mathbf T_{\mathrm{bulk/interface}}
=
-\int_{\partial B_k(t)} (\mathbf x-\mathbf X_k)\times \Pi_f\mathbf n_k\, \text{d}S
+\int_{B_k(t)} (\mathbf x-\mathbf X_k)\times \nabla\Pi_k\, \text{d}V.
\end{equation}
Using again the divergence theorem,
\begin{equation}
\int_{B_k(t)} (\mathbf x-\mathbf X_k)\times \nabla\Pi_k\, \text{d}V
=
\int_{\partial B_k(t)} (\mathbf x-\mathbf X_k)\times \Pi_k\mathbf n_k\, \text{d}S,
\end{equation}
hence
\begin{equation}
\Delta\mathbf T_{\mathrm{bulk/interface}}
=
\int_{\partial B_k(t)} (\mathbf x-\mathbf X_k)\times q_k\mathbf n_k\, \text{d}S.
\end{equation}
The singular surface term contributes
\begin{equation}
\Delta\mathbf T_{\mathrm{sing}}
=
-\int_{\partial B_k(t)} (\mathbf x-\mathbf X_k)\times q_k\mathbf n_k\, \text{d}S,
\end{equation}
so again
\begin{equation}
\Delta\mathbf T_k=0.
\end{equation}

Therefore the transformation
\begin{equation}
p\mapsto p^\ast=p+\Pi,
\qquad
\mathbf f\mapsto \mathbf f^\ast=\mathbf f+\nabla_{\mathcal D}\Pi
\end{equation}
leaves unchanged the velocity field, the net force and torque on every rigid body, and their translational and rotational motions. It only shifts the isotropic stress by $-\Pi\mathbf I$. This construction is specific to incompressible media. In a compressible material, $p$ is tied to the constitutive state, so replacing $p$ by $p+\Pi$ generally changes the motion. In an incompressible medium, however, the pressure field is a constraint reaction rather than a constitutive state variable: it is determined only insofar as needed to enforce $\nabla \cdot \mathbf{v} = 0$.

The transformation also has the chain property: applying first $\Pi_1$ and then $\Pi_2$ gives
\begin{equation}
(p+\Pi_1+\Pi_2,\; \mathbf f+\nabla_{\mathcal D}(\Pi_1+\Pi_2)).
\end{equation}
So the argument remains valid even when the starting force density already contains singular interfacial terms from a previous transformation. This derivation shows that the total electromagnetic force on objects in incompressible fluids can be directly computed from the volume integral of \cref{eq:f_ng} only.

\subsection{Proof 4: Minkowski stress tensor in lossy media}

Here, we prove \cref{eq:minkowski_to_prove}. We expand the different terms of time-averaged Minkowski stress tensor.
First, for the electric part,
\begin{equation}
    \frac{1}{2} \nabla \cdot \Re[\varepsilon \widetilde{\mathbf{E}} \otimes \widetilde{\mathbf{E}}^*] = \frac{1}{2} \Re[(\nabla \cdot (\varepsilon \widetilde{\mathbf{E}}))\widetilde{\mathbf{E}}^* + \varepsilon(\widetilde{\mathbf{E}} \cdot \nabla)\widetilde{\mathbf{E}}^*] = \frac{1}{2} \Re[\rho_f\widetilde{\mathbf{E}}^*] + \frac{1}{2} \Re[ \varepsilon(\widetilde{\mathbf{E}} \cdot \nabla) \widetilde{\mathbf{E}}^*]
\end{equation}
and
\begin{equation} \label{eq:e_squared_term}
\begin{split}
     - \frac{1}{4} \Re[\nabla(\varepsilon |\widetilde{\mathbf{E}}|^2)] &= - \frac{1}{4} \Re[\varepsilon \nabla |\widetilde{\mathbf{E}}|^2+|\widetilde{\mathbf{E}}|^2\nabla \varepsilon] \\ &= - \frac{1}{4} \Re[|\widetilde{\mathbf{E}}|^2\nabla \varepsilon + 2 \varepsilon \Re[(\widetilde{\mathbf{E}}^*\cdot\nabla)\widetilde{\mathbf{E}}+ \widetilde{\mathbf{E}}^*\times (\nabla \times \widetilde{\mathbf{E}})]] \\ &= - \frac{1}{4} |\widetilde{\mathbf{E}}|^2\nabla\Re\varepsilon - \frac{1}{2} \Re\varepsilon  \Re[(\widetilde{\mathbf{E}}^*\cdot\nabla)\widetilde{\mathbf{E}}] - \frac{1}{2} \Re\varepsilon \Re[\widetilde{\mathbf{E}}^*\times (\nabla \times \widetilde{\mathbf{E}})].
\end{split}
\end{equation}
Summing the two,
\begin{equation}
    \frac{1}{2} \nabla \cdot \Re[\varepsilon \widetilde{\mathbf{E}} \otimes \widetilde{\mathbf{E}}^*] - \frac{1}{4} \Re[\nabla(\varepsilon |\widetilde{\mathbf{E}}|^2)] = \frac{1}{2} \Re[\rho_f\widetilde{\mathbf{E}}^*] - \frac{1}{4} |\widetilde{\mathbf{E}}|^2\nabla\Re\varepsilon + \frac{1}{2} \Im\varepsilon \Im[(\widetilde{\mathbf{E}}^*\cdot\nabla)\widetilde{\mathbf{E}}] - \frac{1}{2} \Re\varepsilon \Re[\widetilde{\mathbf{E}}^*\times (\nabla \times \widetilde{\mathbf{E}})].
    \label{eq:electro_part}
\end{equation}
Similarly, for the magnetic part,
\begin{equation}
    \frac{1}{2} \nabla \cdot \Re[\mu \widetilde{\mathbf{H}} \otimes \widetilde{\mathbf{H}}^*] - \frac{1}{4} \Re[\nabla(\mu |\widetilde{\mathbf{H}}|^2)] = - \frac{1}{4} |\widetilde{\mathbf{H}}|^2\nabla \Re\mu + \frac{1}{2} \Im\mu \Im[(\widetilde{\mathbf{H}}^*\cdot\nabla)\widetilde{\mathbf{H}}] - \frac{1}{2} \Re\mu \Re[\widetilde{\mathbf{H}}^*\times (\nabla \times \widetilde{\mathbf{H}})].
    \label{eq:magneto_part}
\end{equation}
Finally, using Ampère and Faraday's laws, it is possible to show that
\begin{equation}
\begin{aligned}
        - \frac{1}{2} \Re\varepsilon \Re[\widetilde{\mathbf{E}}^*\times &(\nabla \times \widetilde{\mathbf{E}})] - \frac{1}{2} \Re\mu \Re[\widetilde{\mathbf{H}}^*\times (\nabla \times \widetilde{\mathbf{H}})] \\ &\quad = \frac{1}{2} \Re[\widetilde{\mathbf{J}}_f\times \widetilde{\mathbf{B}}^*] + \frac{1}{2} \Im\varepsilon \Im[\widetilde{\mathbf{E}}^*\times (\nabla \times \widetilde{\mathbf{E}})] +  \frac{1}{2} \Im\mu \Im[\widetilde{\mathbf{H}}^*\times (\nabla \times \widetilde{\mathbf{H}})].
\end{aligned}
\end{equation}
Summing \cref{eq:electro_part,eq:magneto_part}, and rewriting the free charge part with the help of \cref{eq:free_charge} proves the stated result.

\newpage
\bibliographystyle{unsrtnat}
\bibliography{supp}